

\font\titlefont = cmr10 scaled\magstep 4
\font\sectionfont = cmr10
\font\littlefont = cmr5
\font\eightrm = cmr8

\def\ss{\scriptstyle}
\def\sss{\scriptscriptstyle}

\newcount\tcflag
\tcflag = 0  

\ifnum\tcflag = 0 \magnification = 1200 \fi  

\global\baselineskip = 1.2\baselineskip
\global\parskip = 4pt plus 0.3pt
\global\abovedisplayskip = 18pt plus3pt minus9pt
\global\belowdisplayskip = 18pt plus3pt minus9pt
\global\abovedisplayshortskip = 6pt plus3pt
\global\belowdisplayshortskip = 6pt plus3pt

\def\barsoff{\overfullrule=0pt}


\def\endignore{}
\def\ignore #1\endignore{}

\newcount\dflag
\dflag = 0


\def\monthname{\ifcase\month
\or January \or February \or March \or April \or May \or June%
\or July \or August \or September \or October \or November %
\or December
\fi}

\newcount\dummy
\newcount\minute  
\newcount\hour
\newcount\localtime
\newcount\localday
\localtime = \time
\localday = \day

\def\advanceclock#1#2{ 
\dummy = #1
\multiply\dummy by 60
\advance\dummy by #2
\advance\localtime by \dummy
\ifnum\localtime > 1440 
\advance\localtime by -1440
\advance\localday by 1
\fi}

\def\settime{{\dummy = \localtime%
\divide\dummy by 60%
\hour = \dummy
\minute = \localtime%
\multiply\dummy by 60%
\advance\minute by -\dummy
\ifnum\minute < 10
\xdef\spacer{0} 
\else \xdef\spacer{}
\fi %
\ifnum\hour < 12
\xdef\ampm{a.m.} 
\else
\xdef\ampm{p.m.} 
\advance\hour by -12 %
\fi %
\ifnum\hour = 0 \hour = 12 \fi
\xdef\timestring{\number\hour : \spacer \number\minute%
\thinspace \ampm}}}



\def\endtitle{}
\def\title#1\endtitle{\vskip.5in\titlefont
\global\baselineskip = 2\baselineskip
#1\vskip.4in
\baselineskip = 0.5\baselineskip\rm}

\def\endauthors{}
\def\authors#1\endauthors{#1}

\def\endabstract{}
\def\abstract#1\endabstract{\vskip .3in%
\centerline{\sectionfont\bf Abstract}%
\vskip .1in
\noindent#1}

\def\nopageonenumber{\footline={\ifnum\pageno<2\hfil\else
\hss\tenrm\folio\hss\fi}}  

\newcount\nsection
\newcount\nsubsection

\def\section#1{\global\advance\nsection by 1
\nsubsection=0
\bigskip\noindent\centerline{\sectionfont \bf \number\nsection.\ #1}
\bigskip\rm\nobreak}

\def\subsection#1{\global\advance\nsubsection by 1
\bigskip\noindent\sectionfont \sl \number\nsection.\number\nsubsection)\
#1\bigskip\rm\nobreak}

\def\topic#1{{\medskip\noindent $\bullet$ \it #1:}}
\def\endtopic{\medskip}

\def\appendix#1#2{\bigskip\noindent%
\centerline{\sectionfont \bf Appendix #1.\ #2}
\bigskip\rm\nobreak}


\newcount\nref
\global\nref = 1

\def\ref#1#2{\xdef #1{[\number\nref]}
\ifnum\nref = 1\global\xdef\therefs{\item{[\number\nref]} #2\ }
\else
\global\xdef\oldrefs{\therefs}
\global\xdef\therefs{\oldrefs\vskip.1in\item{[\number\nref]} #2\ }%
\fi%
\global\advance\nref by 1
}

\def\listrefs{\vfill\eject
\centerline{\sectionfont \bf References}
\bigskip\rm\nobreak \therefs}


\newcount\nfoot
\global\nfoot = 1

\def\foot#1#2{\xdef #1{(\number\nfoot)}
\footnote{${}^{\number\nfoot}$}{\eightrm #2}
\global\advance\nfoot by 1
}


\newcount\nfig
\global\nfig = 1

\def\fig#1{\xdef #1{(\number\nfig)}
\global\advance\nfig by 1
}


\newcount\cflag
\newcount\nequation
\global\nequation = 1
\def\eqlabel{(1)}

\def\nexteqno{
\global\advance\nequation by 1
\xdef\eqlabel{(\number\nequation)}}

\def\lasteqno{\global\advance\nequation by -1
\xdef\eqlabel{(\number\nequation)}}

\def\label#1{\xdef #1{(\number\nequation)}
\ifnum\dflag = 1
{\escapechar = -1
\xdef\draftname{\littlefont\string#1}}
\fi}

\def\clabel#1#2{\xdef\eqlabel{(#2 \number\nequation )}
\global\cflag = 1
\xdef #1{\eqlabel}
\ifnum\dflag = 1
{\escapechar = -1
\xdef\draftname{\string#1}}
\fi}

\def\cclabel#1#2{\xdef\eqlabel{#2)}
\global\cflag = 1
\xdef #1{\eqlabel}
\ifnum\dflag = 1
{\escapechar = -1
\xdef\draftname{\string#1}}
\fi}


\def\eeq{}

\def\eqnn #1\eeq{$$ #1 $$}

\def\eq #1\eeq{
\ifnum\dflag = 0
{\xdef\draftname{\ }}
\fi 
$$ #1
\eqno{\eqlabel \rlap{\ \draftname}} $$
\nexteqno}







\def\eqa #1\eeq{
\ifnum\dflag = 0
{\xdef\draftname{\ }}
\fi 
$$ \eqalignno{ #1 } $$
\global\cflag = 0}


\def\ie{{\it i.e.\/}}
\def\eg{{\it e.g.\/}}
\def\etc{{\it etc.\/}}

\def\apriori{{\it a priori\/}}

\def\adhoc{{\it ad hoc\/}}


\def\mpla#1#2#3{{\it Mod.\ Phys.\ Lett.} {\bf A#1}, (19#2) #3}

\def\npb#1#2#3{{\it Nucl.\ Phys.} {\bf B#1} (19#2) #3}
\def\plb#1#2#3{{\it Phys.\ Lett.} {\bf #1B} (19#2) #3}

\def\prd#1#2#3{{\it Phys.\ Rev.} {\bf D#1} (19#2) #3}

\def\prep#1#2#3{{\it Phys.\ Rep.} {\bf C#1} (19#2) #3}
\def\prl#1#2#3{{\it Phys.\ Rev.\ Lett.} {\bf #1} (19#2) #3}


\global\nulldelimiterspace = 0pt



\def\frac#1#2{{{#1} \over {#2}}\,}  
\def\hf{{1\over 2}}
\def\nth#1{{1\over #1}}


\def\Square{{\vbox {\hrule height 0.6pt\hbox{\vrule width 0.6pt\hskip 3pt
        \vbox{\vskip 6pt}\hskip 3pt \vrule width 0.6pt}\hrule height 0.6pt}}}
\def\Dsl{\hbox{/\kern-.6700em\it D}} 
\def\dsl{\hbox{/\kern-.5300em$\partial$}}
\def\pxpsl{\hbox{/\kern-.5600em$p$}}
\def\ssl{\hbox{/\kern-.5300em$s$}}
\def\epssl{\hbox{/\kern-.5100em$\epsilon$}}
\def\delsl{\hbox{/\kern-.6300em$\nabla$}}
\def\lxpsl{\hbox{/\kern-.4300em$l$}}
\def\elxpsl{\hbox{/\kern-.4500em$\ell$}}
\def\kxpsl{\hbox{/\kern-.5100em$k$}}
\def\qxpsl{\hbox{/\kern-.5000em$q$}}
\def\sla#1{\raise.15ex\hbox{$/$}\kern-.57em #1}



\def\twi{\widetilde}

\def\roughly#1{\mathrel{\raise.3ex\hbox{$#1$\kern-.75em\lower1ex\hbox{$\sim$}}}}

\def\ol#1{\overline{#1}}





\def\Scd{{\cal D}}

\def\Scg{{\cal G}}

\def\Sci{{\cal I}}

\def\Scl{{\cal L}}

\def\Sco{{\cal O}}

\def\Scw{{\cal W}}


\def\ssa{{\sss A}}

\def\ssd{{\sss D}}

\def\ssf{{\sss F}}

\def\ssi{{\sss I}}
\def\ssj{{\sss J}}

\def\ssl{{\sss L}}

\def\ssn{{\sss N}}

\def\ssv{{\sss V}}
\def\ssw{{\sss W}}

\def\ssy{{\sss Y}}


\def\Tr{\mathop{\rm Tr}}

\def\Re{{\rm Re\;}}
\def\Im{{\rm Im\;}}



\def\avg#1{\langle #1 \rangle}

\def\Avg#1{\left\langle #1 \right\rangle}



\def\cc{{\rm c.c.}}


\nopageonenumber
\baselineskip = 18pt
\barsoff


\def\Kahler{K\"ahler}
\def\expval{{\it vev}}
\def\Lc{\Lambda_c}
\def\bk{\item{}}
\def\ov#1{#1^*}
\def\d{\delta}

\def\WWn{W_n^\alpha W_{\alpha n}}
\def\WW{W^\alpha W_\alpha}
\def\lbarl{\lambda^\alpha \lambda_\alpha}
\def\Dterm#1{\left( #1 \right)_\ssd}
\def\Fterm#1{\left( #1 \right)_\ssf}

\def\mp{M_{\sss P}}

\def\Sbar{S^*}
\def\S0bar{S_0^*}
\def\Dbar{\overline{D}}
\def\DDbar{\overline{DD}}
\def\Ubar{U^*}
\def\Uhat{\widehat{U}}
\def\Wh{{\cal W}}

\def\W{(W_{np})}
\def\cu{{,u}}

\def\cs{{,s}}

\def\IZ{{\bf Z}}
\def\Sckk{K_{np}}


\rightline{CERN-TH/95-111}
\rightline{hep-th/9505171}

\title
\centerline{On Gaugino Condensation with}
\centerline{Field-Dependent Gauge Couplings}
\endtitle

\authors
\centerline{{\bf C.P. Burgess,${}^a$ J.-P. Derendinger,${}^b$
F. Quevedo${}^b$ and M. Quir\'os ${}^c$}}
\vskip .07in
\centerline{\it ${}^a$ Physics Department, McGill University}
\centerline{\it 3600 University St., Montr\'eal, Qu\'ebec,
Canada, H3A 2T8.}
\vskip .03in
\centerline{\it ${}^b$ Institut de Physique, Universit\'e de Neuch\^atel}
\centerline{\it CH-2000 Neuch\^atel, Switzerland.}
\vskip .03in
\centerline{\it ${}^c$ Theory Division, CERN, CH-1211 Geneva 23,
Switzerland}
\endauthors

\vskip .1in
\vskip .3in
\vbox{\baselineskip 11pt
We study in detail gaugino condensation in globally and locally supersymmetric
Yang-Mills theories. We focus on models for which gauge-neutral matter couples
to the gauge bosons only through nonminimal gauge kinetic terms, for the cases
of one and several condensing gauge groups. Using only symmetry arguments, the
low-energy expansion, and general properties of supersymmetry, we compute the
low energy Wilson action, as well as the (2PI) effective action for the
composite {\it classical} superfield $U\equiv\langle \Tr\WW \rangle$,
with $W_\alpha$ the supersymmetric gauge field strength. The 2PI effective
action provides a firmer foundation for the approach of Veneziano and
Yankielowicz, who treated the composite superfield, $U$, as a quantum degree
of freedom. We show how to rederive the Wilson action by minimizing
the 2PI action with respect to $U$. We determine, in both formulations and for
global and local supersymmetry, the effective superpotential, $W$, the
non-perturbative contributions to the low-energy K\"ahler potential $K$, and
the
leading higher supercovariant derivative terms in an expansion in inverse
powers
of the condensation scale. As an application of our results we include the
string
moduli dependence of the super- and K\"ahler potentials for simple
orbifold models.}


\vskip .15in
\leftline{CERN-TH/95-111, IEM-FT-105/95, McGill-95/05, NEIP-95--04}
\leftline{May 1995}


\vfill\eject
\section{Introduction}

\ref\dynamicalbreaking{E. Witten, \npb{188}{81}{513};
\npb{202}{82}{253}.}
\ref\AKMRV{For a review, see D. Amati, K. Konishi, Y. Meurice,
G. C. Rossi and G. Veneziano, {\it Phys. Rep.} {\bf 162} (1988) 169.}
\ref\VY{
G. Veneziano and S. Yankielowicz, \plb{113}{82}{231};\bk
T. R. Taylor, G. Veneziano and S. Yankielowicz, \npb{218}{83}{493}.}
\ref\ADS{I. Affleck, M. Dine and N. Seiberg, \prl{51}{84}{1026};
\npb{241}{84}{493}.}
\ref\FGN{H.P. Nilles, \plb{115}{82}{193};\bk
S. Ferrara, L. Girardello and H. P. Nilles, \plb{125}{83}{457}.}
\ref\DIN{
J.-P. Derendinger, L. E. Ib\'a\~nez and H. P. Nilles,
\plb{155}{85}{65}.}
\ref\DRSW{
M. Dine, R. Rohm, N. Seiberg and E. Witten, \plb{156}{85}{55}.}
\ref\Taylor{T. R. Taylor, \plb{164}{85}{43}.}
\ref\KP{
C. Kounnas and M. Porrati, \plb{191}{87}{91}.}

\ref\FILQ{
A. Font, L. E. Ib\'a\~nez, D. L\"ust and F. Quevedo, \plb{245}{90}{401};\bk
S. Ferrara, N. Magnoli, T. R. Taylor and G. Veneziano, \plb{245}{90}{409};\bk
H.P. Nilles and M. Olechowski, \plb{248}{90}{268};\bk
P. Bin\'etruy and M.K. Gaillard, \plb{253}{91}{119}.}
\ref\DMR{A. de la Macorra and G.G. Ross, \npb{404}{93}{321};\bk
R. Peschanski and C.A.  Savoy, Saclay Preprint, SPhT-95/002, 1995.}

Understanding just how supersymmetry breaks is the biggest
present uncertainty for formulating realistic supersymmetric field
theories. It is also the principal obstacle to obtaining
realistic string models. A tantalizingly attractive possibility is
for supersymmetry to be broken dynamically
\dynamicalbreaking,  such as by the condensation of an
asymptotically-free gauge group \AKMRV,\VY,\ADS. A particularly
appealing version of this theme uses the condensation of the gauginos
in such a gauge theory, and this option has been explored in
many settings, including global supersymmetry \AKMRV,
supergravity \FGN\ and string theory \DIN,\DRSW,\Taylor,\KP,\FILQ.
Unfortunately, although there is agreement on the behaviour of
the simplest systems, the non-perturbative nature of this mechanism
has led to a certain amount of disarray in the literature, with
several different, not necessarily equivalent, approaches having
emerged towards finding the low-energy predictions of these
theories, well below the  condensation scale \VY,\FGN,\DRSW,\DMR.

In this article we re-examine the gaugino condensation
process yet one more time. One naturally ventures onto
such well-trodden territory with trepidation, however we
do so here with two goals in mind. Our first goal is
to clarify the relationship amongst the results of the original workers,
and to do so in a way which makes clear which of their conclusions
are robust. In particular, we show how the main two alternative analyses
of gaugino condensation can be thought of as equivalent approaches that
respectively  formulate the problem in terms of the `Wilson', and the
`effective' actions (see below for the precise definitions) for the
theory. Our second goal is to take advantage of the systematic nature of our
approach to see how the standard  results can be extended. Besides
reproducing the usual expression for the superpotential for the low-energy
matter fields, we find also a new result: the contributions to the
\Kahler\ potential due to the strongly-interacting gauge physics.
Our methods can also be used to find further subleading corrections to these
in powers of supercovariant derivatives divided by the condensation scale.

The rest of this article starts with a short preamble, followed by
the three sections, which contain our three main results. The
preamble, which makes up the contents of the next section, consists
of a whirlwind review of the effective and Wilson actions for a
generic field theory. This is introductory material on which the rest of
our arguments are based. The guts of our results are presented in section 3,
where we derive in detail the effective and Wilson actions for
gaugino condensation within global supersymmetry. We confine our analysis
to a pure super-Yang-Mills theory coupled to gauge-neutral matter through
nonminimal gauge kinetic terms. (We do not pursue here the possibility of
additional  matter supermultiplets within the strongly-coupled gauge
sector.) We keep our treatment completely general, by identifying the most
general possible action of each kind that is consistent with the
symmetries of the problem, and with the low-energy expansion.

The main purpose of section 3's analysis is twofold. The first purpose
is to make contact with the earlier literature. Our treatment of the
superpotential in the Wilson action duplicates that of Dine {\it et al.}
\DRSW, deriving the standard form using the anomalous $R$ invariance
of the underlying theory. The same analysis, when applied to the
2PI effective action for the gaugino bilinear field, reproduces the results
of Veneziano and Yankielowicz (VY) \VY, although with a few important
differences. We therefore provide a more solid basis for their approach, as
well as show how it can be generalized to include other effects, such as
subleading corrections in $\Lc/M$, the influence of more complicated
field-dependent couplings, or more complicated gauge groups. We also show
how to re-derive the Wilson action from the 2PI effective action, and
thereby clarify the relation between these two approaches.

\ref\KL{V. Kaplunovsky and J. Louis, \npb{422}{94}{57}.}
\ref\LNN{Z. Lalak, A. Niemeyer and H.P. Nilles, Munich preprint
TUM-HEP-211/95.}
\ref\shore{G.M. Shore, \npb{222}{83}{446}.}

Even though the 2PI effective action reproduces the VY results, there are a
number of important differences between their approach and ours which are
instructive to identify. First, at a technical level, the treatment using the
2PI formalism permits us to show quite generally that the effective action must
be {\it linear} in the gauge-coupling field, $S$, something that is an \adhoc\
assumption in earlier discussions \VY,\KL. Second, although we appeal
to the same symmetries as in Ref.~\VY, our approaches differ
fundamentally in how these symmetries are implemented. In the spirit of
anomaly matching, VY write an effective action for which the anomalies
in the underlying theory are expressed by the non-invariance of the
effective action. In particular,  based on a physical motivation, they choose
one particular noninvariant form for this purpose, but this choice is not
unique. The uncertainty in the argument therefore enters in quantifying the
validity of the assumed form for the low-energy non-invariance. By
contrast, our own approach follows the old `spurion' ideas by identifying an
exact symmetry of the underlying theory by cancelling the anomaly against
the classical transformation of some of the low-energy `external' fields
(a similar method was also used in Ref.~\KL\ ).
The requirement that the low-energy theory be {\it invariant} with respect
to this symmetry then dictates the form for the 2PI superpotential, as well
as the low-energy limit of the \Kahler\ potential. We therefore avoid the
ambiguity of the earlier approach.

A more conceptual difference between our analysis and that
of VY comes from the role that is played by the composite gaugino
condensate field, $U$, in the two approaches. In the VY approach, this field
is a {\it bona-fide} quantum field --- much like the $\eta$ particle in QCD
--- which must be integrated out to obtain the Wilson action for
energies well below the condensation scale
(see \KL, \LNN\ for a recent discussion). In our 2PI approach,
however, $U$ arises purely as a classical field which represents the
expectation value of a particular composite field on which we have chosen
to focus  (see also \shore\ for a similar treatment). As a result, the
Wilson action for the low-energy fields is obtained
simply by evaluating the effective action at its stationary point, with no
further path integration over strongly-coupled degrees of freedom
required at all.

The second, and final, purpose of this section (still section 3) is to proceed
beyond the standard results. To do so we use the constraints which follow
from the theory's anomalous $R$ symmetry and scale invariance to determine the
first subleading terms in the low-energy action for the light fields. We are
led in this way to the nonperturbative contributions to the \Kahler\ potential
of the low-energy theory. We identify additional corrections, which arise as an
expansion in powers of supercovariant derivatives of the light fields, divided
by the condensation scale.

Moving on to section 4, we generalize the previous arguments to supergravity,
using the formalism of superconformal fields. The principal new complication is
that the scale and $R$ symmetries of the global case are now part of the full,
local superconformal symmetry, whose anomalies  {\it must} be cancelled by a
Green-Schwarz counterterm. Even though the formalism is completely different
from the global case since, in particular, we  have to work with a compensating
field  (as usual in conformal supergravity), we arrive at similar results,
both for the Wilson and the 2PI actions, as in the global case. We verify that
the results of global supersymmetry are reobtained from those of supergravity
in
an inverse Planck mass expansion. Within this framework we repeat the
derivation
of the Wilson action from the effective action, and so find the
Planck-mass-suppressed corrections to the low-energy action.

In section 5, we specialize the supergravity results to derive explicit
formulae for the super- and \Kahler-potentials of the low-energy theory, as
functions of the  moduli ($T$) for (2,2) string vacua.
We find in this way explicit results,  for single- and multiple-gauge-group
condensations, for which all of the low-energy symmetries are
manifest, even after including threshold
corrections.

Our conclusions are briefly summarized in section 6. We present in an
appendix, useful formulae concerning the component expression
of the superconformal action, details about fixing the compensating
field to obtain standard Poincar\'e supergravity as well as
the derivation of its correspondence with the globally supersymmetric limit.

\section{The Effective vs Wilson Actions}

The two main tools for our analysis of gaugino condensation are the theory's
Wilson and effective actions. Each of these quantities contains a specific
kind of information about the nature of the theory.  Since much of
our discussion hinges on the properties of, and the relationship between,
these two kinds of effective actions, we pause here to outline their
definitions and differences.

\ref\cornwall{J.M. Cornwall, R. Jackiw and E. Tomboulis,
\prd{10}{74}{2428}. }
\ref\peskin{M. Peskin, in {\it Recent Advances in
Field Theory and Statistical Mechanics}, Les Houches 1982, ed. by J.-B.
Zuber and R. Stora (North Holland, Amsterdam, 1984).}

In quantum field theory the generating functional of one-particle-irreducible
(1PI) correlation functions --- the `effective action' --- is the
relevant quantity with which to ask questions related to the
expectation values of the field  operators in the vacuum state (\expval's).
This is because the \expval\ of these fields must minimize the effective
action. It is therefore the most useful tool for analysing issues of symmetry
breaking. A similar, two-particle-irreducible (2PI) generalization is
of interest to determine the \expval\ of composite field bilinears \cornwall,
\peskin.

\ref\Russians{M.A. Shifman and A.I. Vainshtein, \npb{277}{86}{456};
\npb{359}{91}{571}.}

On the other hand, `the' Wilson action  is obtained by integrating out all of
those modes whose masses lie above some reference energy
scale.\foot\whyquotes{We  put the word `the' here in quotes, since there are as
many  kinds of Wilson actions as there are ways of distinguishing the low from
the high energies within a theory.} This provides a  useful way of organizing
the relative effects for low-energy observables of physics  associated with
much
higher scales. In particular,  for gaugino condensation it encapsulates the
effective  description of physics well below the condensation scale, after all
composites of the strongly interacting fields are integrated out. The Wilson
action is a  local functional of the `light' quantum fields. In supersymmetric
theories it is the Wilson action which has many useful properties, such as the
requirement that some interactions must depend only  holomorphically on chiral
superfields.\foot\russia{For a discussion, see for instance ref. \Russians.}

\subsection{The Effective Action}

We start with the definition of the effective action. Consider a field theory,
whose fields we generically call $\phi$. In order to determine the vacuum
expectation value of an operator $\Sco[\phi]$, we couple an external current
to it and compute the  generating functional for its connected Green
functions as follows \foot\start{In ref.~\cornwall\ the composite
operators $\ss \Sco(\phi)$ are taken in such a way that each of the fields
$\ss \phi$ were evaluated at different spacetime points and coupled to a
non-local current $\ss J(x_1, x_2)$. This generalization is not relevant for
our
purposes of finding the vacuum state and so we consider the full
composite operator at a single point $\ss x$. }
\label\eqone
\eq
\exp \Bigl\{i \Wh[J] \Bigr\} = \int \Scd\phi \; \exp\left\{ i \int d^4x
\; \Bigl[ \Scl[\phi] + J \Sco[\phi] \Bigr] \right\}.
\eeq

We define the effective action for the operator $\Sco$ by performing
a Legendre transformation on the functional $\Wh[J]$. That is, we
define the average field, $u$, by:
\label\eqtwo
\eq
u(J) \equiv { \delta \Wh \over \delta J} = \Avg{\Sco[\phi]}_\ssj,
\eeq
where the average in this last equation denotes the quantity
\label\eqthree
\eq
\Avg{\Sco[\phi]}_\ssj \equiv e^{-i\Wh[J]} \; \int \Scd\phi \; \Sco[\phi] \;
\exp\left\{ i \int d^4x \; \Bigl[ \Scl[\phi] + J \Sco[\phi] \Bigr] \right\} .
\eeq
The Legendre transform of $\Wh[J]$ is then the functional $\Gamma[u]$,
defined by
\label\eqfour
\eq
\Gamma[u] \equiv \Wh[J(u)] - \int d^4x \; u \, J ,
\eeq
where we imagine $J(u)$ to be the external current that is required
to obtain the expectation value $\Avg{\Sco}_\ssj = u$, and which may
be found, in principle, by inverting eq.~\eqtwo\ for $u(J)$.  If $\Gamma[u]$
is known, $J(u)$ can be found by directly differentiating the
defining equation for $\Gamma[u]$:
\label\eqfive
\eq
{\delta \Gamma[u] \over \delta u} + J = 0.
\eeq

A path-integral expression for $\Gamma[u]$ may be obtained by combining
the definitions of eqs.~\eqone\ and \eqfour:
\label\eqsix
\eq
\exp \Bigl\{i \Gamma[u] \Bigr\} = \int \Scd\phi \; \exp\left\{ i \int d^4x
\; \Bigl[ \Scl[\phi] + J \left( \Sco[\phi] - u \right) \Bigr] \right\} .
\eeq
It is important to keep in mind that this equation is somewhat
self-referential since the current $J$ that appears on the right-hand side is
meant to be that function of $u$ given by $J = - (\delta \Gamma / \delta u)$.
This does not make the above equation useless for computing $\Gamma[u]$.
On the contrary, this choice for the current merely ensures that, in
perturbation theory,  the appropriate reducible graphs get cancelled, making
$\Gamma[u]$  the sum of a set of irreducible graphs. For example, if
$\Sco(\phi) = \phi$, then $\Gamma$ is simply the sum of all one-particle
irreducible (1PI)  graphs. If $\Sco(\phi) = \phi^2$, then it is the sum of
two-particle  irreducible (2PI) graphs \cornwall, and so on.

\ref\gammaasenergy{T.D. Lee and G.C. Wick, Phys. Rev. {\bf D9} (1974) 2291;\bk
S. Coleman, {\it Laws of Hadronic Matter}, edited by A. Zichichi (Periodici
Scientifici, Milan, 1975);\bk
M. Peskin, ref.~\peskin;\bk
C.P. Burgess and G. Kunstatter, \mpla{2}{87}{875}.}

For our purposes, the most important property of the functional
$\Gamma[u]$, is that its stationary point specifies the \expval\ of the
original operator, $\Sco(\phi)$. This can be seen from eqs.~\eqtwo\ and
\eqfive\ above. Eq.~\eqtwo\ shows that for an arbitrary current, $J$, $u$
gives the value of $\Avg{\Sco(\phi)}$ in the presence of this current. But we
are really interested in this expectation in the absence of all external
currents, and by eq.~\eqfive\ this corresponds to choosing $u$ to satisfy
$\delta \Gamma / \delta u = 0$. For time-independent field configurations
this argument can be sharpened to show that $\Gamma[u]$ is the minimum
expectation value of the system's Hamiltonian given that the expectation of
the field $\Sco(\phi)$ is constrained to equal $u$ \gammaasenergy.

\subsection{The Wilson Action}

It is often the case that we are only concerned with the properties of very
low-energy phenomena in the field theory of interest. For example, we might
imagine only being interested in the effective action, $\Gamma[u]$, for
fields $u$ which vary only over distances that are much longer than $\ell =
1/\mu$. (For example, we might demand, in a Euclidean-space formulation,
that only Fourier components of $u$ for which the four-momentum,
$p_\alpha$, satisfies $p^2\le \mu^2$  are of interest.) We can ensure that
such a condition is true for $u(x)$ by demanding the same of the external
current, $J(x)$, with which we probe  the system. In this case, we can
partition the integration over $\phi$ into an integration over modes
$\phi_\ell(\mu)$ and $\phi_h(\mu)$, where the modes  $\phi_\ell(\mu)$
satisfy  $p^2 \le \mu^2$, and the modes $\phi_h(\mu)$  do not. Typically
there are many ways to define this split between high and low energies, and
there is in principle a different Wilson action for each. This `scheme
dependence' does not concern us in detail here, although we assume in later
sections that it is possible to define this split in a manifestly
supersymmetric way.

The above construction is particularly simple when $\Sco(\phi) = \phi$
(for which we use the notation $\varphi$ in place of $u$), because in this
case the integration over $\phi_h$ completely decouples from the  external
current,  allowing us to write eq.~\eqsix\ in the following, suggestive,
form:
\label\eqseven
\eq
\exp \Bigl\{i \Gamma[\varphi] \Bigr\} = \int \Scd\phi_\ell \; \exp\left\{ i
\int d^4x \; \Bigl[ \Scl_\ssw[\phi_\ell,\mu] + J \left( \phi_\ell -
\varphi \right) \Bigr] \right\} . \eeq
Here $\Scl_\ssw$ is the Lagrangian density for the Wilson action,
$I_\ssw[\phi_\ell,\mu] \equiv \int d^4x \;  \Scl_\ssw(\phi_\ell,\mu)$,
which is defined by
\label\eqeight
\eq
\exp \Bigl\{i I_\ssw[\phi_\ell,\mu] \Bigr\} \equiv \int \Scd\phi_h \;
\exp\left\{ i \int d^4x \;  \Scl[\phi_\ell,\phi_h]  \right\} .
\eeq
By writing the Wilson action in terms of a Lagrangian density, we
anticipate one of its most important properties: that it is a local
functional of the fields once it is written as a power series expansion in
$1/\mu$. The same need not be true, in general, of the effective action,
$\Gamma[u]$, since this is defined to include the integration over {\it all}
modes of the underlying fields, including any massless modes from which
potentially nonlocal contributions can come.

There is an important situation for which the effective and Wilson
actions are very simply related to one another. If we evaluate the effective
action at its stationary point: $\Gamma[\ol{u}]$, with $(\delta \Gamma/
\delta u)_{u = \ol{u}} = 0$, then the corresponding current vanishes,
$J(\ol{u}) = 0$. In this case we have
\label\equivalence
\eq
\exp \Bigl\{i \Gamma[\ol{u}] \Bigr\} = \int \Scd\phi_\ell \ \exp\left\{ i
\int d^4x \; \Scl_\ssw[\phi_\ell,\mu]  \right\} .
\eeq
Integrating over the light modes $\phi_\ell$ corresponds formally to
the $\mu\rightarrow 0$ limit.
This expression states that the result obtained by completely integrating
out {\it all} of the modes of the field gives the same answer as would be
obtained by evaluating the effective action at its minimum. This equality
should be thought of as an equivalence in the dependence of both sides on
whatever other parameters (background fields, \etc) may characterize the
system.

The above connection between the Wilson and effective actions has concrete
applications to systems for which an entire sector of the theory is much
more massive than the scale $\mu$ which defines the Wilson action. In this
case it is useful to define the Wilson action in such a way that all
fields from this sector are completely integrated out. If $A$
collectively denotes these massive fields, and $\phi$ denotes the light
fields whose masses are lighter than $\mu$, then the Wilson action for
$\phi$ can be written as:
\label\wilsonresult
\eq \eqalign{
\exp \Bigl\{i I_\ssw[\phi_\ell] \Bigr\} &= \int \Scd\phi_h \,\Scd A \,
\exp\left\{ i \int d^4x \; \Scl[\phi_\ell,\phi_h,A]  \right\}  \cr
& = \int \Scd \phi_h \, \exp \Bigl\{i \Gamma_\ssa[\phi_\ell, \phi_h]
\Bigr\} ,\cr
\hbox{where} \qquad  \exp \Bigl\{i \Gamma_\ssa[\phi_\ell,\phi_h] \Bigr\}
&= \int \Scd A \, \exp\left\{ i \int d^4x \; \Scl[ \phi_\ell, \phi_h,A]
\right\}.  \cr}
\eeq
$\Gamma_\ssa[\phi]$ is the result obtained after the complete integration
only over the fields $A$. The other fields, $\phi_\ell$ and $\phi_h$, can be
considered to be simply background fields for this part of the integration.
Keeping in mind eq.~\equivalence, it can therefore equally well be regarded
as the result obtained by evaluating the effective action for some operator
involving only the fields $A$, evaluated at its minimum --- again with
$\phi_\ell$ and $\phi_h$ regarded as fixed background fields. The full
Wilson action at scale $\mu$ is then simply found by integrating this result
over $\phi_h$. It is in this vein that we use the effective action in this
paper, where the heavy sector is made up of the strongly-coupled gauge
theory.

\section{Gaugino Condensation in Global Supersymmetry}

This section is devoted to applying the above definitions to
determine the low-energy Wilson action for a collection of
gauge-singlet chiral matter superfields --- in components: $\Sigma_\ssi
= \{z_\ssi, \psi_\ssi, f_\ssi \}$ with $I = 1, \dots,N$ --- well below
the condensation scale, $\Lc$, for some nonabelian gauge theory, having
gauge group $\Scg$ (\eg\ the hidden  $E_8$ sector of $(2,2)$
compactifications). Although  we work with chiral superfields, we certainly
do not exclude the possibility of there also being gauge interactions
[including the standard $SU_c(3) \times SU_\ssl(2) \times U_\ssy(1)$] in
the low-energy theory below $\Lc$, so long as these are weakly coupled at
scales  $\Lc$ and higher. In general, the $\Sigma_\ssi$ can carry gauge
quantum numbers for these other gauge interactions, whose effects at $\Lc$
can be computed perturbatively.  Since this low-energy gauge dynamics
plays no role in what follows, for simplicity we ignore it from here on.

We next write down our starting action, $\Sci$. We do not restrict
ourselves to renormalizable couplings above the condensation scale, $\Lc$,
since we regard our initial classical action to be itself a Wilson action
obtained by integrating out more physics --- perhaps string physics --- at
still higher scales, $M \gg \Lc$. We suppose that this action is
supersymmetric, and in the present section we also assume that $M$ is
sufficiently small in comparison to the Planck scale, $\mp$,  to justify a
globally supersymmetric treatment. (Relaxing this last assumption is the
topic of section 4, below). To leading order in $1/M$ we must neglect all
terms which are suppressed by any inverse powers of $M$. As is typical
for supersymmetric theories, since we do {\it not} know \apriori\ whether
the scalar fields, $z_\ssi$, have \expval's which are small compared
to $M$, we do {\it not} assume these to be small and so do {\it not} expand
$\Sci$ in powers of $\Sigma_\ssi$. It is therefore convenient  to scale the
appropriate power of $M$ out of the $z_\ssi$ to ensure that they, and
so also the $\Sigma_\ssi$, are  dimensionless. We {\it do}, however, assume
all supersymmetry-breaking \expval's to be much smaller than $M$, and so
we {\it do} neglect all higher supercovariant derivatives of the various
fields.

We first consider the case of a strongly-interacting simple gauge sector.
The leading terms in  $1/M$ then are\foot\dfformula{
We use the standard notation $(\ldots)_F= \int d^2\theta\,(\ldots)$ and
$(\ldots)_D = \int d^2\theta d^2\overline\theta\,(\ldots)$.}:
\label\actionon
\eq
\Sci = \int d^4x \; \left[ \Dterm{ K_p(\Sigma_\ssi ,\Sigma_\ssj^*)} +\left[
\Fterm{ W_p( \Sigma_\ssi ) +  {1\over4}f_p(\Sigma_\ssi)\, \Tr \WW } +
\cc\right]\right],
\eeq
where $K_p$, $W_p$ and $f_p$ are arbitrary functions of their arguments,
and where $W_p$ and $f_p$ both must be holomorphic functions. The
subscript `$p$' on these functions could stand for `prior' (or, for string
theory, `perturbative', since in the usual string scenarios these functions
can be computed  in string perturbation theory without any recourse to
strong coupling).  $W_\alpha \equiv -{1\over 4} \Dbar \Dbar\,( e^{-V}
D_\alpha \, e^V)$ is the usual left-handed chiral spinor
superfield which contains the gauge field strengths, that is constructed
from the gauge-potential superfield $V = V^a T_a$, with $V^a= \{
\lambda^a_\ssl, A^a_\mu, D^a \}$ in the Wess-Zumino gauge. (The
generators, $T_a$, used in $V$ are normalized by the condition
$\Tr(T_a T_b) =
\delta_{ab}$.) We assume that the matrix elements of the components of
$W_\alpha$ are also much smaller than $M$, and so, in writing
eq.~\actionon, we have also neglected all higher powers of this
gauge-field-strength superfield.

Notice that the only coupling between the gauge and matter sectors at this
point is through the nonminimal gauge-kinetic term,
$\Fterm{{1\over4}f_p(\Sigma_\ssi)
\Tr \WW }$, and so it is on this term that we now focus.  It is convenient to
follow string notation and define the gauge coupling  $f_p(\Sigma_\ssi)$ as
one of the chiral  fields of the problem, which we denote hereafter as
$S = \{s,\psi_s, f_s\}$, the others being generically referred to as $\Sigma$.
Provided that $f_p(\Sigma_\ssi)$ is not a constant, this
can always be achieved by performing a suitable holomorphic field
redefinition amongst the $\Sigma_\ssi$'s. To leading order in $1/M$ the
implications of the strong gauge dynamics for the light fields,
$\Sigma_\ssi$, are therefore completely encapsulated in the integral of this
term with respect to the gauge configurations, $V$:
\label\whatwewant
\eq
\exp\Bigl\{ i \Gamma_\ssv[S] \Bigr\} = \int \Scd V \exp
\Bigl\{ i I[S,V] \Bigr\} ,
\eeq
with
\label\eqten
\eq \eqalign{
I[S,V] &\equiv {1\over4}\int d^4x \; \left[ \Fterm{ S \Tr\WW } + \cc \right]
\cr
&= \int d^4x \; \left\{ \Re s \; \left[ \hf \, D^a D^a - \nth{4} \,
F^a_{\mu\nu}
F^{a\mu\nu} +{i\over2} \, \ol{\lambda}^a \Dsl \lambda^a \right]
+ \nth{4} \, \Im s \; F^a_{\mu\nu} \twi{F}^{a\mu\nu} +\cdots\right\}. \cr}
\eeq
We omit in this equation the terms involving $\psi_s$ and $f_s$ as well as the
coupling of $\Im s$ to the derivative of the chiral gaugino current.
Notice that for the special case of a spatially constant, supersymmetric,
scalar field configuration, $\avg{S}  \equiv \{ \avg{s}, 0,0 \}$, the constant
$\avg{s}$ is related to the gauge coupling constant, $g$, and vacuum angle,
$\Theta$, according to:
\label\eqtenptone
\eq
\avg{s} = {1 \over g^2} + {i \Theta \over 8 \pi^2}.
\eeq
\ref\susymatter{N. Seiberg, \plb{318}{93}{469};
\prd{49}{94}{6857}; preprint RU-94-64, IASSNS-HEP-94/57,
hep-th/9408013.}

It is the purpose of the remainder of this section to determine the result of
the integration over the  gauge multiplet, $V$. In order to do so we make one
standard, yet crucial, dynamical assumption. We assume that all of the
gauge nonsinglet particles become confined into bound
states whose masses are of the  order of the gauge theory's condensation
scale, $\Lc$.  It is important in what follows that there be no very light
bound states with masses that are sufficiently small to appear in the
low-energy theory. This
assumption is consistent with the symmetries of the pure supersymmetric
Yang-Mills system. The same need not be true once strongly-coupled matter
multiplets are also considered into the analysis, and so this assumption
precludes their inclusion (unless they happen to be free of accidental
symmetries which can keep their bound states systematically light). In more
general cases we see no difficulty in extending the analysis along the lines
of refs.~\susymatter, \VY, \AKMRV, \KL.

We follow two approaches to writing down $\Gamma_\ssv[S]$. In the first
approach we imagine $\Gamma_\ssv[S]$ to be the result
obtained when successively integrating out lower and lower
frequency modes of the gauge sector {\it \`a la} Wilson.  We then
recompute the same result using a 2PI effective  action for the gaugino
bilinear field, $\Tr \WW$. We reproduce in this second way the results of
ref.~\VY, although we are able to do so based on what we think are much
more firmly-grounded assumptions, in which the nature of the
approximations being made are more easily seen.

\subsection{The Wilson Approach}

The total effective action can be written as
\label\marianoa
\eq
\Gamma[S,\Sigma]=\Gamma_p[S,\Sigma]+\Gamma_V[S]
\eeq
where
\label\marianob
\eq
\Gamma_p[S,\Sigma]=\int d^4x \left[ \Dterm{K_p(S,S^*,\Sigma,\Sigma^*)}
+\left[ \Fterm{W_p(\Sigma)}+\cc\right]\right]
\eeq

We wish to constrain the general form that is permitted for
$\Gamma_\ssv[S]$, as defined in eq.~\whatwewant:
\label\eqeleven
\eqnn
\exp \Bigl\{ i\Gamma_\ssv[S] \Bigr\} = \int \Scd V \; \exp \left\{ i \int
d^4x \; \left[ \Fterm{ {1\over4}S \Tr \WW } + \cc \right]  \right\}.
\eeq
Imagine that we do so by first considering the functional integration of the
gauge multiplet only down to a lower scale, $M'$, from the initial scale, $M$,
at which the original action is as in eq.~\eqten. The result obtained
reproduces $\Gamma_\ssv[S]$ once $M'$ is taken to zero.  For $M'$ larger
than $\Lc$, however, the action obtained in this way depends on the gauge
potential, $V$, in addition to the background field, $S$.

Denote the result of such an integration by $I_\ssw[S,V]$: it is the Wilson
action defined at scale $M'$. It enjoys
the following two very important properties:  ($i$) {\it locality}, and
($ii$) {\it supersymmetry}. Imagine expanding the action,
$I_\ssw[S,V] = \int d^4x  \, \Scl_\ssw$, in powers of the
supercovariant derivatives, $D_\alpha S$, of the field, $S$. On dimensional
grounds, each power should appear premultiplied by an inverse power of
$M^{1/2}$ or $(M')^{1/2}$. Supersymmetry and locality dictate that the
lowest-dimension
terms may be written:
\label\eqfourteenptone
\eq \eqalign{
\Scl_\ssw &= \left[\Fterm{W_{np}(S)}  + \cc \right] +  \Dterm{K_{np}(S,S^*)}
+   \Dterm{H_{np}(S,S^*) \; (DDS)  + \cc } \cr
& \qquad + \left[{1\over4}\Fterm{ F(S) \Tr \WW} + \cc \right]  + \cdots \cr}
\eeq
and, according to \marianoa, this leads to the total super and K\"ahler
potentials:
\eq\eqalign{
W(S,\Sigma) & = W_p(\Sigma)+W_{np}(S) \cr
K(S,S^*,\Sigma,\Sigma^*) & = K_p(S,S^*,\Sigma,\Sigma^*)
+ K_{np}(S,S^*) \cr   }
\eeq
The subscript `$np$' refers now to `non-prior', which again, for the
superpotential will also mean non-perturbative given that all its perturbative
corrections vanish. As is well known, the same is not true for the \Kahler\
potential, which does receive perturbative contributions, as we discuss
below. The ellipses in  this expression represent terms
involving four or more supercovariant derivatives. Notice that supercovariant
derivatives need not be considered in the $F$-term above, since any such
contribution can be rewritten as a $D$-term using identities like
$(S^m[\overline{DD}\overline S]^n)_F = -4(S^m \overline S [\overline{DD}
\overline S]^{n-1})_D$.  We now turn to a discussion of the form that can be
taken by each of these terms.

\topic{The Superpotential}

We start with some standard arguments which constrain the superpotential,
$W(S)$.  We exploit the classical $R$ symmetry of the lagrangian of
eq.~\eqten, under which all fermion fields, $\lambda^a$ and
$\psi_s$, rotate by a common phase $e^{i\beta}$:
\label\eqthirteen
\eq \eqalign{
W_\alpha (x,\theta) &\to e^{i\beta} \;  W_\alpha\left(x,
e^{-i\beta}\theta\right) \cr
S(x,\theta) &\to S \left(x, e^{-i\beta}\theta\right) \cr} .
\eeq
Although this symmetry is anomalous, its anomaly can be cancelled by
supplementing the transformation law for $S$ of eq.~\eqthirteen\ with:
\label\eqfourteen
\eq
S \to S - {4ib\over 3} \; \beta,
\eeq
where $b$ is a known constant to be calculated below.

\ref\WDS{E. Witten, \npb{268}{86}{79};\bk M. Dine and N. Seiberg,
\prl{57}{86}{2625}.}

Nontrivial information can now be obtained provided that one is
prepared to assume that the above transformations, eqs.~\eqthirteen\ and
\eqfourteen, are exact symmetries of the Wilson theory. This is equivalent to
assuming the Adler-Bardeen theorem for the $R$-symmetry anomaly, \ie\ that the
anomaly cancellation is automatically exact once it has been enforced to one
loop. If so, then $W_{np}(S)$ must be an $R$-invariant superpotential, which is
to say that $W_{np} \to e^{2i\beta} W_{np}$ \DRSW, (see also \KP\ ). This
implies that the superpotential must be given by:
\label\eqfourteenptfour
\eq
W_{np}(S) = w \exp\left[-\ {3 S \over 2 b} \right] ,
\eeq
where $w$ is a constant\foot\nonren{A way to derive the
nonrenormalization theorem for the superpotential in
 perturbation theory \WDS\
is to realize that the symmetry \eqthirteen, \eqfourteen\
leads necessarily to a superpotential as in \eqfourteenptfour\
but with $\ss w=0$ because the gauge coupling is related to
$\ss \avg{S}$
by $\ss g^2= (\avg{\Re S})^{-1}$.
Perturbation theory is an expansion in negative powers of
$\ss \avg{\Re S}$
and therefore $\ss \exp\left[-\ {3 S \over 2 b} \right]$ can only
 have nonperturbative
origin.}
 which we may take on dimensional grounds
to be proportional to $M^3$.
This expression for the superpotential, $W_{np}(S)$, is familiar
from the literature.

\topic{The Gauge Kinetic Function}

A similar result constrains the nonminimal gauge kinetic terms in
eq.~\eqfourteenptone\ \Russians:
\label\eqfifteen
\eq
{1\over4} \left[ \Fterm{ F(S) \Tr \WW } + \cc \right] ,
\eeq
where the new coefficient, $S' \equiv S(M') \equiv F(S)$, is some function of
the original quantity, $S=S(M)$, as well as of the two scales, $M$ and $M'$.
This function can be considered to define a scheme for the running of the
gauge coupling due to the formula, eq.~\eqtenptone, which relates $\Re s$ to
the gauge coupling, $1/g^2$ defined at scale $M$. According to this scheme,
we identify the (Wilson) gauge coupling, $g'$, at scale $M'$, using the same
relation:  $\Re s' = 1/(g')^2$.

The supersymmetry of eq.~\eqfifteen\ requires
that the expression $S' = S'(S,M/M')$ must be holomorphic in the original
variable $S$. (As is argued in detail in ref.~\Russians, this holomorphy is
justified because we are focussing on the couplings of the Wilson action,
rather than, say, the 1PI effective action.)  But this condition of holomorphy
implies that any dependence of $S'$ on $\Re s = 1/g^2$ is intimately related
to its dependence on $\Im s = \Theta/(2\pi^2)$, about which we know a
great deal. In particular, because we know that no dependence whatsoever on
$\Theta$ can be generated within perturbation theory, we know that $\Re S'$
must be completely independent of $\Im S$ in
perturbation theory. The same conclusion follows as an exact result if
$F(S)$ is required to be invariant under the $R$-transformation of
eq.~\eqfourteen.

These conditions have as their only solution
\label\eqsixteen
\eq \eqalign{
S' &= S + B\left( {M \over M'} \right) \cr
&= S - b \, \log\left( {M^2 \over M'^2} \right) + O\left( {M' \over M} \right),
\cr}
\eeq
where we have used the fact that the \apriori\ arbitrary function $B(M/M')$
must arise independently of $S$, and so can be computed purely at one loop.
Eq.~\eqsixteen\ shows that the running of the gauge coupling, $g$, as
defined by the evolution of the supersymmetric Wilson action, only gets a
contribution at one loop \Russians:
\label\RGeqtn
\eq
{1 \over g'^2} = {1 \over g^2} - b \, \log\left( {M^2 \over M'^2} \right).
\eeq
The constant $b$ that appears here is the same as the constant which
appears in eq.~\eqfourteen, and is recognized as the one-loop beta-function
coefficient for a supersymmetric theory. It is therefore explicitly given by:
\label\eqeighteen
\eq
b =  {1 \over 16 \pi^2 } \left[  3 C(G) - \sum_\ssi T\left(R_\ssi \right)
\right], \eeq
where  the quantity
$T(R)$ is the index for the representation $R$  of the gauge
group $G$, $C(G)$ denotes its quadratic Casimir [$C(G)=T(Adj\,\,G)$] and
$R_\ssi$ represents the representation of whatever matter multiplets
the theory may contain.

For future purposes, the exact running of $g$ given in eq.~\RGeqtn\ permits
the definition of a renormalization group (RG)-invariant scale,
$\Lambda$, defined by:
\label\LcDef
\eq
\Lambda = M \exp\left( -\; {1\over 2b g^2} \right) = M \exp\left( -\; {(s+s^*)
\over 4b} \right) .
\eeq
We expect the condensation scale, $\Lc$, to be of the same order of
magnitude as $\Lambda$.

\vfill\eject

\topic{The \Kahler\ Potential}

We next turn to the general constraints on the form for the \Kahler\
potential, $K_{np}(S,S^*)$. The anomaly-free combination of $R$ and shift
symmetries used above implies the exact result that $K_{np}$ cannot depend
on $\Im S$: $K_{np} = K_{np}(S+S^*)$. (This same result alternatively follows
to all orders in perturbation theory from its independence from $\Theta$ in
the perturbative approximation.)

The unknown function, $K_{np}(S+S^*)$, must
satisfy one further constraint, which expresses the independence of all
physical results on the floating scale $M$. To implement this condition
it is important to remember how $M$ drops out of low-energy results. There
are two ways in which this happens. Some of the $M$-dependence simply
cancels the explicit $M$ dependence which is already present in $K_p$.
$K_p$ depends on $M$ as well as on the various physical mass parameters
(call them $m_i$) of the high-energy theory. For string theory, for example,
the $m_i$ would denote the masses of the various multiplets that are
heavier than $M$ and so have been integrated out to obtain the Wilson action
at the floating scale $M$. After the cancellation of the $M$ dependence in
$K_p$, the physical content of the higher-energy physics that $K_p$ encodes
is set by the physical masses, $m_i$. For the present purposes we regard
this higher-energy physics to be known, and so we focus here on the
contribution
of the strongly-coupled gauge sector at lower energies.

The second way for $M$ to cancel out of physical results is for explicit
$M$ dependence to cancel the $M$-dependence that is implicit in the coupling,
$S+S^* = 1/g^2$. This $M$-dependence of the coupling is as
is required by the RG equation, as expressed by eq.~\RGeqtn. We therefore
write:
\label\ksplit
\eq
K_{np} = K_{np-{\rm ct}}(S+S^*,M) + K_{np-{\rm inv}}(S+S^*,M,M'),
\eeq
where $K_{np-{\rm ct}}$ is defined by the condition
\label\defcondition
\eq
K_p(S+S^*,M,m_i) + K_{np-{\rm ct}}(S+S^*,M) = \hbox{independent of $M$}.
\eeq
$K_{np-{\rm inv}}$, on the other hand, contains the contribution due to the
strongly-coupled gauge physics, and must be RG invariant. That is, on
dimensional grounds $K_{np-{\rm inv}}$ may always be written:
\label\dimgrounds
\eq
K_{np-{\rm inv}}(S+S^*) = M'^2 \Scg\left(S+S^*,{M\over M'} \right),
\eeq
where RG invariance implies that the unknown dimensionless function,
$\Scg(x,y)$, satisfies:
\label\RGDE
\eq
{d \Scg \over d M} = 0 = {d (S+S^*) \over d M} \; { \partial \Scg
\over \partial x} + {1\over M'} \; {\partial \Scg \over \partial y} \; .
\eeq
This last equation has as its general solution
\label\RGsoln
\eq
K_{np-{\rm inv}} = M'^2 \; k \left( {\Lambda \over M'} \right) .
\eeq
$\Lambda$ here denotes the RG-invariant scale as defined by eq.~\LcDef.
Thus general considerations determine the running \Kahler\ potential,
$K_{np-{\rm inv}}$, in terms of a function, $k(z)$, of the single variable,
$z=\Lambda/M'$.

More can be inferred by considering various limiting cases. The perturbative
limit is given by $M' \gg \Lambda$, or $z \ll 1$. In this limit we expect
$K_{np-{\rm inv}}$ to admit an expansion in powers of $g^2$ or $g'^2$:
\label\PertK
\eq  \eqalign{
k(z) &= \sum_n k_n \; g'^{2n} \cr
&= \sum_n k_n \left({-1 \over 2b \log z }
\right)^n . \cr}
\eeq

The other limit of interest is $z \to \infty$, since this corresponds to $M'
\to
0$, with $\Lambda$ fixed. This is the \Kahler\ potential which results when
the gauge sector has been integrated out completely. In this case the limiting
form follows from the assumption that the spectrum of the gauge sector
contains no very light states. As a result the Wilson action, and $K_{np-{\rm
inv}}$ in
particular, should not become singular as $M' \to 0$. This implies that the
asymptotic form for $k(z)$ for large $z$ must be $k \sim k_\infty z^2 +
(\hbox{subleading terms})$, where $k_\infty$ is a (possibly zero) constant.
The resulting \Kahler\ potential then is:
\label\Kahlerresult
\eq  \eqalign{
K_{np-{\rm inv}}(S,S^*) &= \lim_{M' \to 0} M'^2 \; k\left( {\Lambda \over M'}
\right) \cr
&= k_\infty \; \Lambda^2 \cr
&= k_\infty \; M^2 \exp\left[ - \; {(S+S^*) \over 2b} \right] ;\cr}
\eeq
a result which is in any case clear on dimensional grounds.

This last expression gives the non-perturbative part of the \Kahler\ potential
that is produced by integrating out the strongly-coupled gauge physics at the
condensation scale.

\topic{Higher-Derivative Terms}

The unknown function, $H_{np}(S,S^*)$, which premultiplies the leading
higher-derivative term can be determined using the same arguments as
were used above to fix $K_{np}$. As before we can separate and discard
those $M$-dependent pieces --- \ie\ $H_{np-{\rm ct}}$ --- which serve
to cancel the $M$-dependence of any higher-derivative terms appearing
in the higher-energy theory. We focus the remainder of this section on
determining the remaining term, $H_{np-{\rm inv}}$.
In this case, since the $R$ symmetry
rotates the derivatives according to $DD S \to
e^{2i\beta} \;  DD S$, we find:
\label\Hresult
\eq
H_{np-{\rm inv}}(S,S^*) = M' \exp\left[ {3 (S-S^*) \over 4 b} \right] \;
h\left(
{\Lambda \over M'} \right).
\eeq
$h(z)$ is an unspecified function of the RG-invariant variable $z=
\Lambda/M'$, which is itself a function of the $R$-invariant quantity
$S+S^*$ [see eq. \LcDef].

A nonsingular result as $M' \to 0$ in this case implies the
asymptotic form $h(z) \sim h_\infty z + (\hbox{subleading terms})$,
and  the result for $H_{np-{\rm inv}}$ in this limit therefore is:
\label\HforMzero
\eq \eqalign{
H_{np-{\rm inv}}(S,S^*) &= h_\infty \Lambda \; \exp\left[ {3 (S-S^*) \over 4 b}
\right]
\cr &= h_\infty M \; \exp\left[ {1 \over 2 b} (S - 2 S^*) \right] . \cr}
\eeq

Notice that,  as expected, these higher-derivative terms are suppressed by
powers of the condensation scale, $\Lc \sim \Lambda$, rather than by $M$.
This makes it legitimate to work with these higher-derivative corrections
while continuing to ignore the higher-derivative terms in the initial
action, $\Sci$, at scale $M$. Higher-derivative corrections to the
effective lagrangian, such as these, appear not to have been considered
before in the literature. Clearly their neglect is justified to the extent that
the supersymmetry-breaking \expval's, such as the auxiliary field $f_s$, are
much smaller than the condensation scale.

\endtopic

These results can be combined to obtain the Wilson action for the full
low-energy theory --- neglecting powers of $1/M$ but not necessarily
powers of $1/\Lc$ --- by ($i$) adding the nonperturbative piece just found
(for $M' \to 0$) to $K_p(S,S^*)$ and $W_p(S)$ of the high-energy action, and
($ii$) integrating the result over the high-frequency part of $S$. That is, if
$K_{\rm tot} = K_p  + K_{np} + O \left( {1\over M} \right)$ and $W_{\rm tot}
= W_p + W_{np} + O \left( {1\over M} \right)$, then
\label\hereswilson
\eq
\exp\Bigl\{ i I_\ssw[ \Sigma_\ssi,\mu ] \Bigr\} = \int \Scd
(\Sigma_\ssi)_h \exp \left\{ i \int d^4x \Bigl[ \Dterm{ K_{\rm tot} } +
\Fterm{ W_{\rm tot} } \Bigr] + \cdots \right\} ,
\eeq
where the ellipses denote terms involving higher supercovariant derivatives,
such as in eq.~\eqfourteenptone.

\subsection{Using the 2PI Action}

To obtain more detailed information as to how the strong gauge dynamics
generates the quantity $\Gamma_\ssv[S]$ just found, we next use the 2PI action
to compute the expectation value of the gaugino bilinear, $\Avg{\Tr \lbarl}$.
We
wish to show that this approach reproduces our earlier result for
$\Gamma_\ssv[S]$.  We also wish to make contact with, and improve on, the
results of earlier workers, starting with ref.~\VY.

\ref\GGRS{S.J. Gates, M.T. Grisaru, M. Ro\v cek and W. Siegel,
{\it Superspace}, Benjamin-Cummings (1983).}
\ref\BDQQ{C. P. Burgess, J.-P. Derendinger, F. Quevedo and M. Quir\'os,
\plb{B348}{95}{428}.}
\ref\BGT{P. Binetruy, M.K. Gaillard and T.R. Taylor, preprint
LBL-36744-UCB-95/03-95 (1995).}
We start by coupling an external chiral supermultiplet of currents,
$J$, to the chiral scalar superfield, $\Tr \WW$, which contains as its
lowest component the gaugino bilinear, $\Tr \lbarl$:
\label\eqtwentytwo
\eq
\exp \Bigl\{i {\cal W}_{np} [J,S] \Bigr\} = \int \Scd V \; \exp \left\{ i \int
d^4x \;
{1\over4}
\left[ \Fterm{ S \Tr \WW } + \Fterm{J \Tr \WW} + \cc \right]  \right\}.
\eeq
In terms of the chiral Legendre transformed
superfield:\foot\chiralderiv{The
definition of the functional derivative with respect to a chiral superfield
($\ss {\d \Phi(x')\over \d \Phi(x)} = \DDbar\d(x-x')$) is given, for instance,
in ref.~\GGRS.}
\label\eqtwentythree
\eq
U \equiv 4\,{\delta {\cal W}_{np} \over \delta J} = \Avg{\Tr \WW}_\ssj,
\eeq
 the corresponding 2PI effective action for $U$ becomes:
\label\eqtwentyfour
\eq
\Gamma_{np}[U,S] \equiv {\cal W}_{np} [J,S] - {1\over4}
\int d^4x \left[ \Fterm{U \, J} + \cc \right].
\eeq

Using now \marianob, the total effective action can be decomposed as:
\label\marianod
\eq
\Gamma[S,U,\Sigma]=\Gamma_p[S,\Sigma]+\Gamma_{np}[S,U]
\eeq
and, accordingly, the total super and K\"ahler potentials as:
\label\marianof
\eq\eqalign{
W(S,U,\Sigma) & = W_p(\Sigma)+W_{np}(S,U) \cr
K(S,S^*,U,U^*,\Sigma,\Sigma^*) & = K_p(S,S^*,\Sigma,\Sigma^*)+
K_{np}(S,S^*,U,U^*) \cr  }
\eeq
As usual, the variables $J$ and $U$, satisfy the relation:
\label\eqtwentyfive
\eq
{\delta\over\delta U}\,\Gamma_{np}[U,S] + {1\over4}J = 0,
\eeq
and so the physical case of vanishing external current corresponds to
choosing $U$ to lie at a stationary point of $\Gamma_{np}[U,S]$.

An important property of $\Gamma_{np}[U,S]$ follows from the fact that the
generating functional, ${\cal W}_{np}[J,S]$, only depends on its two arguments,
$J$
and $S$, through their sum: ${\cal W}_{np}[J,S] = {\cal W}_{np}[J+S]$, as may
be seen from
eq.~\eqtwentytwo. This turns out to imply the following {\it exact} property
for the 2PI function, $\Gamma_{np}[U,S]$:
\label\eqtwentysix
\eq
{ \delta \over \delta S}\,\Gamma_{np}[U,S]  = {\delta\over\delta J}\,{\cal
W}_{np}[J,S]
= {1\over4}U.
\eeq
for all $S$ and $U$. As a consequence, the $S$-dependence of $\Gamma_{np}[U,S]$
is completely determined to be
\label\eqtwentyseven
\eq
\Gamma_{np}[U,S] = \Xi[U] + {1\over4}
\int d^4x \left[ \Fterm{U \, S} + \cc \right].
\eeq

In principle, since the functional $\Xi[U]$ represents an effective (as
opposed to Wilson) action, it need not be a local functional of $U$.
However, it is here that our dynamical assumption --- that  the spectrum of
the gauge sector involve only bound states of masses  equal to or larger than
the condensation scale, $\Lc$ --- plays a role. So long as we consider only
configurations, $U$, which vary appreciably over very long distances compared
to $1/\Lc$, $\Xi[U]$ can be taken to be a local expansion in powers of
the supercovariant  derivatives $D^\alpha$ of the fields $U$
\foot\shoreagain{See Ref.~ \shore\ for a discussion on
higher supercovariant derivatives of $U$.} and $S$.
This is the domain of interest for determining the vacuum
expectation value for $U$. Notice also that since $\Xi[U]$ does not depend on
$S$, the scale against which its higher-derivative dependence must be
compared is $M$ rather than $\Lc$. This is because no factors of $\Re S$
can arise to convert powers of $M$ into powers of $\Lc$. Since we work only
to leading order in $1/M$, we therefore neglect all supercovariant
derivatives in $\Xi[U]$ in what follows. $\Xi[U]$ can therefore be written
\label\eqtwentyeight
\eq
\Xi[U] = \int d^4x \left[ \Dterm{ \Sckk (U,\Ubar)} + \left(
\Fterm{F(U) } + \cc \right) \right].
\eeq

We next determine the implications of the accidental $R$
symmetry for $\Xi[U]$. In terms of the variables $U$ and $S$, the
symmetry of eqs.~\eqthirteen\ and \eqfourteen\ of the path
integral becomes
\label\eqtwentynine
\eq
U \to e^{2i\beta} U \qquad \hbox{and} \qquad S \to S - {4 i b \over 3} \;
\beta,
\eeq
together with the corresponding rotation of the coordinates,
$\theta_\ssl$. This symmetry dictates that $\Sckk(U,U^*)$ must be
a function only of the invariant combination $\Ubar U$, and that
$F(U)$ must take the form:

\label\eqthirty
\eq
F(U) = {b \over 6} \; U \log\left( {\zeta U \over M^3} \right),
\eeq
where $\zeta$ is an arbitrary dimensionless constant, which we  expect to be
$O(1)$. This form for the superpotential was first obtained in ref.~\VY.

We pause to remark at this point that this symmetry argument leading to the
superpotential \eqthirty\ is actually a statement of anomaly cancellation. To
see this notice that the chiral superfield $U$ satisfies a constraint which
follows from the equation
\label\CS
\eq
\Tr W^\alpha W_\alpha = \overline{DD} \Omega, \qquad\qquad
\Tr \overline W_{\dot\alpha}\overline W^{\dot\alpha} =
DD\Omega,
\eeq
which relates the chiral superfield $\Tr W^\alpha W_\alpha$ to the Chern-Simons
superfield $\Omega$, a real vector superfield. It then follows from the
definition of $U$  [see eqs. \eqtwentytwo\ and \eqtwentythree]
that there exists a real vector superfield $V$ such that
\label\UV
\eq
U = \overline{DD} V , \qquad\qquad
\overline U = DD V,
\eeq
$V$ being defined up to the addition of a linear multiplet \BDQQ.
\foot\binn{This fact has also been observed in Ref.~
\BGT.} In terms of
components, eq. \UV\ has a unique consequence: the imaginary part of the
highest
component of $U$ is a total derivative,
\eq
\Im f_u = \partial_\mu v^\mu.
\eeq
Now, under the symmetry of eq.~\eqtwentynine, the term ${1\over4}(US)_F+\cc$
appearing in $\Gamma_{np}[U,S]$ transforms into a total derivative:
${2b\over3}\beta(\partial^\mu v_\mu)$, and it is this anomaly of the term
${1\over4}(US)_F$ that is cancelled by the transformation of $\Xi[U]$.

Notice that the entire nonperturbative effective superpotential $W_{np}$
can be combined into the form
\label\eqthirtyone
\eq  \eqalign{ W_{np}(U,S)=
{F(U)} + {1\over4}{U \; S} &= {b \over 6} \; U \; \log\left( { \zeta
U e^{3S/2b} \over  M^3} \right) \cr
 &\equiv {b \over 6} \; U \; \left[ \log\left( {U \over  \Lc^3}
\right) - 1 \right], \cr}
\eeq
where we use this last equality to precisely define the condensation scale:
$\Lc(S) = \left( M/\zeta^{{1 \over 3}} \right) \exp\left[-(S/2b)
-\nth{3} \right]$.

More can also be said about the \Kahler\ potential, $\Sckk(U,\Ubar)$, which
appears in eq.~\eqtwentyeight. {\it A-priori}, $R$-invariance and
dimensional analysis permit an arbitrary function of the invariant
dimensionless combination $\Ubar U/M^6$:
\label\interkform
\eq
\Sckk(U,\Ubar) = M^2 \; \Sckk\left( {\Ubar U \over M^6} \right).
\eeq

The key feature of this last equation is that the absence of a dependence on
$S$ implies that the scale over which the undetermined function, $\Sckk$, can
vary appreciably is set by $U \sim M^3$,
rather than $U \sim \Lc^3$. Provided
the region $U \sim \Lc^3$ is the one that is of interest --- as is the case for
gaugino condensation --- we need only consider the region which satisfies
$\Ubar U \ll M^6$. But the result for $\Sckk$ should not become singular in
this limit, once we have performed the subtractions which renormalize the
composite operator $\Tr \WW$. Neglecting all inverse powers of $M$, and
keeping in mind the vanishing anomalous dimension of the composite field
$\Tr \WW$ \AKMRV, \KL, the only possible $M$-independent expression for $\Sckk$
becomes
\label\finalkform
\eq
\Sckk(U,\Ubar) = a (\Ubar U)^{{1 \over 3}},
\eeq
with $a$ a constant.

Eqs.~\eqthirtyone\ and \finalkform\ contain the complete
result for $\Gamma_{np}[U,S]$, up to $O(1/M)$ corrections.

\subsection{Reproducing $\Gamma_\ssv[S]$ from $\Gamma_{np}[U,S]$}

{}From the general discussion of section 2, we know that $\Gamma_\ssv[S]$
is obtained by minimizing $\Gamma_{np}[U,S]$ with respect to $U$. Notice that
this is an exact relationship, and it is {\it not} to be regarded as a `tree
approximation' to performing a path integral over $U$. We will first
perform the minimization in components and later on in terms of
superfields to obtain more compact expressions.

In order to see what is implied by this minimization, it is useful to restrict
$\Gamma_{np}[U,S]$ briefly to constant configurations of its scalar components:
$s, f_s, u$ and $f_u$. The Lagrangian density for $\Gamma_{np}[U,S]$ in this
case
reduces to:
\label\auxil
\eq
{\cal L}_{aux}=K_{np\cu \ov{u}} f_u \ov{f_u}
-\Bigl[ f_u \W_\cu + f_s \W_\cs +\cc\Bigr]  ,
\eeq
where $K_{np\cu\ov{u}}$ and $\W_\cu$ respectively denote $\partial^2
{\Sckk} /\partial u \partial u^*$ and $\partial W_{np}/\partial u$, \etc. We
must eliminate $u$ and $f_u$ from this lagrangian using their equations of
motion. We can easily solve for $f_u$, to get
\label\fuis
\eq
f_u^* = K_{np\cu\ov{u}}^{-1}\W_\cu \equiv f^*_u(s,u,u^*).
\eeq
Substituting this back into the lagrangian then gives
\label\auxtwo
\eq
{\cal L}_{aux}= - K_{np\cu \ov{u}}^{-1}\left|\W_\cu\right|^2
- \left( f_s \W_\cs +\cc\right)
\eeq
This expression must now be extremized with respect to variations of $u$.

\ref\LT{D. L\"ust and T. Taylor, \plb{253}{91}{335}.}
\ref\buddies{B. de Carlos, J.A. Casas and C. Mu\~noz,
\npb{399}{93}{623}.}

Notice that if we simply neglect the term $f_s \W_\cs$, the minimum is found by
solving $\W_\cu=0$. This is the procedure which has been commonly followed in
the literature.\foot\seekl{See, for example, \LT, \buddies, \KL\ for a
recent discussion.} The rationale for doing so has been that the condition
$\W_\cu=0$ ensures that the auxiliary field $f_u$ vanishes, which is the
condition that the strongly-coupled sector does not itself break supersymmetry.
We see from above that this condition actually cannot be imposed,
since, in general,  it does
not  minimize the 2PI action.  The best one can do is to look for solutions in
the neighbourhood of $f_s=0$, for which $f_s$ is small, and so minimize
eq.~\auxtwo\ perturbatively in powers of $f_s$. This gives as its solution a
minimum, $u=\widetilde{u}(s,f_s) = \widetilde{u}_0 (1 + \widetilde{u}_1  +
\dots)$, where
$\W_\cu (u = \widetilde{u}_0) = 0$ and so on.  The solution so obtained is:
\label\eqthirtytwo
\eq \eqalign{
\widetilde{u}_0 &=  \Lc^3 = {M^3\over \zeta e} \; e^{-3s/2b} , \cr
\widetilde{u}_0  \widetilde{u}_1 &= {a \over 4 b^2} \;
(\widetilde{u}_0^* \widetilde{u}_0)^{\nth{3}} \;
f^*_s, \qquad\qquad{\rm etc}\ldots \cr}
\eeq
This expression for $\widetilde{u}_0$ represents one of the main results of the
2PI approach, giving as it does the leading approximation to $\Avg{\Tr
\lbarl}$.

We may now return to our original goal, which was to eliminate the
entire supermultiplet, $U$, to obtain $\Gamma_\ssv[S] =
\Gamma_{np}[U=\widetilde{U}(S),S]$. Given the explicit form for $\Gamma[U,S]$
found
earlier, the equation of motion of the superfield $U$ is:
\label\euler
\eq
{a\over 3}\; U^{-{2\over 3}} \, \DDbar\, (U^*)^{1\over 3} =
{\partial W_{np} \over \partial U} .
\eeq
With our analysis of the scalar potential in mind, we look for solutions to
this equation in powers of $D_\alpha U$ about the zeroeth-order solution,
for which $\W_\cu = 0$. That is, we take the solution to eq.~\euler\ of the
form
$U = \widetilde{U} \equiv \widetilde{U}_0(1 + \widetilde{U}_1 + \widetilde{U}_2
+ \dots)$, where:
\label\firstfew
\eq\eqalign{
\widetilde{U}_0 &= {M^3 \over \zeta  e} \; e^{-3 S /2b} , \cr
\widetilde{U}_1 & ={a\over 2bM}\,(\zeta e)^{1/3}\, X,\cr
\widetilde{U}_2 &= {a^2 \over 12 b^2M^2}\,(\zeta e)^{2/3}
\; \Bigl[\DDbar (Y X^*)- {1\over2}X^2] \Bigr], \cr}
\eeq
where $Y \equiv e^{(2S - S^*)/2b }$ and $X = \DDbar Y$. The appearance
of supersymmetric derivatives in $\widetilde U_1$, $\widetilde U_2$, $\ldots$
indicates that nonzero constant values of these chiral superfields break
supersymmetry.

The final expression for $\Gamma_\ssv[S]$ follows from substituting
these results into $\Gamma_{np}[U,S]$. The leading ($\widetilde{U}_0$)
term gives the superpotential contribution to $\Gamma_\ssv[S]$ as:
\label\eqthirtythree
\eq
W_{np}(S,U) \left.
\right|_{\widetilde{U}_0(S)} = -\; {b\over 6}\;
\widetilde{U}_0(S) = -\ b\ {M^3 \over 6\zeta e}
\; e^{-3 S /2b} ,
\eeq
in agreement with our previous expression, eq.~\eqfourteenptfour,
and the leading term in the \Kahler\ potential for $\Gamma_\ssv[S]$:
\label\leadingkahler
\eq
K_{np} = a (\widetilde{U}_0^* \widetilde{U}_0 )^{\nth{3}} ,
\eeq
which again agrees with our previous result, eq.~\Kahlerresult, with
$k_\infty = a (M^3 / \zeta e)^{{2 \over 3}}$. Continuing to include
$\widetilde{U}_1$ and
beyond generates the higher-supercovariant derivative terms in
$\Gamma_\ssv[S]$.  In particular one can easily check that the
function $H_{np}(S,S^*)$ in \HforMzero\ is given by
$H_{np}(S,S^*)=h_\infty M Y^*$, where the field $Y$ is that used
in the definition of $\widetilde{U}_n$ ($n\geq 1$) in
\firstfew.

The above approach should be contrasted with the standard procedure that
has often been used in the literature, for which $U$ is either considered as a
quantum field to be integrated out, or is eliminated using the
supersymmetry-preserving condition $\W_\cu = 0$. By recognizing $U$ as a
classical field in the 2PI construction, we avoid the conceptual problems of
trying to interpret $\Gamma_{np}[U,S]$ as a {\it bona fide} low-energy (Wilson)
effective action. Similarly, although the condition $\W_\cu = 0$ suffices
to determine the superpotential of $\Gamma_\ssv[S]$, it does not
correctly reproduce the \Kahler\ potential, or the higher-derivative
corrections. Since neither of these has been discussed in earlier work, this
discrepancy has not yet arisen in practice.

Notice also that it is perfectly consistent not to impose $\W_\cu =
0$ and yet still claim that the strongly-coupled sector does not itself break
supersymmetry. That is, even though it is true that the deviations from $\W_\cu
= 0$ permit the auxiliary field, $f_u$ to become nonzero, these deviations only
make $f_u$ proportional to $f_s$, and so they leave the burden of supersymmetry
breakdown with the low-energy auxiliary field, $f_s$.  Of course, the actual
minimum of the exponential scalar potential corresponding to the
superpotential of eq.~\eqthirtythree\ (or eq.~\eqfourteenptfour) occurs for
$s\to \infty$, and so $\widetilde{u}=0$.  This is the well known runaway
weak-coupling, supersymmetric solution, and gives the result which is
required from general considerations based on the Witten index
\dynamicalbreaking.

\subsection{More Than One Condensate}

\ref\krasnikov{N.V. Krasnikov, \plb{193}{87}{37};\bk
L. Dixon, in {\it The Rice Meeting}, B. Bonner and
H. Miettinen, eds, World Scientific (1990);\bk
J.A. Casas, Z. Lalak, C. Mu\~noz and G.G. Ross,
\npb{347}{90}{243}.}

It is very easy to generalize the above considerations to the case of
several independent gauge groups, $G = \prod_n (\times G_n)$, each
of which may condense separately. This case has attracted much attention
in the string literature because it provides a possible way of
generating a potential for the dilaton
field which could fix the dilaton at a finite
value  and break supersymmetry \krasnikov, \buddies.
In this case the classical lagrangian is:
\label\eqthirtyfour
\eq
\Scl = \sum_n \left[{1\over4}\Fterm{f_n S \; \Tr \WWn} + \cc \right],
\eeq
where $W_{n\alpha}$ is the field strength multiplet for the $n$'th gauge
group factor, and $f_n$ are independent constants (or functions of other
(moduli) fields besides $S$).

The direct symmetry argument for this lagrangian is more complicated
because the anomaly in all of the independent classical $R$ symmetries ---
one for each group factor --- cannot in general be cancelled by a single shift
in $\Im S$.  The analogue of the anomaly-free symmetry of the previous sections
is, in this case, now only an approximate symmetry of the theory.

We therefore analyze this theory by first considering the more general
case where each gauge factor has its own coupling-constant field, $S_n$,
and write:
\label\eqthirtyfive
\eq
\Scl' = \sum_n \left[ {1\over4}\Fterm{S_n \; \Tr \WWn} + \cc \right],
\eeq
with the limit $S_n \to f_n S$ taken at the end of the discussion.

We may proceed either by directly writing down the expression for
$\Gamma_\ssv[S_n]$ that is permitted by the symmetries, or by
constructing the 2PI action and eliminating the condensate fields, $U_n$. We
follow here this second approach, since it contains more information than
does the first. The generating functional for this theory is then obtained by
coupling an independent current, $J_n$, to each of the bilinears, $\Tr \WWn$:
\label\eqthirtysix
\eq \eqalign{
\exp \Bigl\{ i{\cal W}_{np}[J_1,S_1,\dots] \Bigr\} &= \int \prod_n \Scd V_n
\; \exp \left\{ i \int d^4x \; \left[ \sum_n {1\over4}
\Fterm{ (S_n + J_n) \; \Tr \WWn }
+ \cc \right]  \right\} \cr
& = \exp\left\{ \sum_n i{\cal W}_{np}[J_n,S_n] \right\}. \cr}
\eeq
Clearly the total generating functional is simply the sum of the same
simple-gauge-group contribution for each factor of the gauge group.
As a result the same is true for the effective action:
$\Gamma_{np}[U_1,S_1,\dots]= \sum_n \Gamma_{np}[U_n,S_n]$.
The results of the previous sections may then be directly used for the
expression for each of the factors, $\Gamma_{np}[U_n,S_n]$, giving
\label\eqthirtyseven
\eq\eqalign{
\Gamma_{np}[U_1,\dots,U_\ssn,S] & = \sum_n \int d^4x \; \left[
\Dterm{ a_n (U_n^* U_n)^{1/3} } \right.  \cr
& + \left. \left[{b_n \over 6} \Fterm{ U_n \;
\log \left( { \zeta_n  U_n e^{3 f_n S/2b_n} \over  M^3} \right) }
 + \cc\right] \right]. }
\eeq

The stationary point of this potential for the field $\widetilde{U}_n$ needs
again to be
written as  $\widetilde{U}_n = \widetilde{U}_{n0}(S)(1 +
\widetilde{U}_{n1}(S) +\dots) $ which we can find
perturbatively in the supercovariant derivatives. The
first term is therefore
\label\eqthirtyeight
\eq
\widetilde{U}_{n0}(S) = {M^3  \over \zeta_n e} \;  \exp \left( - \; {3f_n S
\over 2b_n}\right) ,
\eeq
and so the low-energy superpotential and \Kahler\ potential for $S$
become:
\label\eqthirtynine
\eq \eqalign{
W_{np}(S) &=-\ {1\over 6}\sum_n b_n \widetilde{U}_{n0}(S)
= -\ {1\over 6} \sum_n \left( {M^3  \over
\zeta_n e}\right) \;  b_n \;  \exp  \left[ - \; {3 f_n S \over 2b_n}
\right]. \cr
K_{np}(S,S^*) &= \sum_n a_n (\widetilde{U}_{n0}^*
\widetilde{U}_{n0} )^{\nth{3}} =
\sum_n a_n \left( {M^3  \over \zeta_n e}\right)^{{2 \over 3}} \;
\exp\left[  - \; {f_n S \over 2b_n} \; (S+S^*) \right] . \cr}
\eeq
Notice that even if the superpotential \eqthirtynine\ has been previously
used  \krasnikov, to our knowledge this is the first time it actually has been
derived.

\section{Supergravity}

\ref\FGKVP{S. Ferrara, L. Girardello, T. Kugo and A. Van Proeyen,
\npb{223}{83}{191}.}
\ref\CFGVP{E. Cremmer, S. Ferrara, L. Girardello
and A. van Proeyen, \plb{116}{82}{231}; \npb{212}{83}{413}.}
\ref\KU{T. Kugo and  S. Uehara, \npb{222}{83}{125};
\npb{226}{83}{49}.}
\ref\L{J. Louis, in Particles, Strings and Cosmology 1991, ed. by
P. Nath and S. Reucroft (World Scientific, Singapore, 1992).}
\ref\DFKZa{J.-P. Derendinger, S. Ferrara, C. Kounnas and
F. Zwirner \npb{372}{92}{145}.}
\ref\DFKZb{J.-P. Derendinger, S. Ferrara, C. Kounnas and
F. Zwirner, \plb{271}{91}{307}.}
\ref\FPvN{S. Ferrara and P. van Nieuwenhuizen, \plb{78}{78}{573}.}
\ref\PvN{P. van Nieuwenhuizen, \prep{68}{81}{189}, section 4.4.}

In this section we extend the previous results to the case of local
supersymmetry, as is necessary if the scale $M$ should be as large as
$\mp$, such as for string theory. For our purposes, we believe that
the most convenient formulation of supergravity coupled to matter
takes advantage of the simplicity of
the superconformal tensor calculus. Poincar\'e supergravity is then
obtained
from superconformal gravity by imposing symmetry-fixing conditions
on certain components of a supermultiplet used as a compensator.
We use the simplest choice of compensator, a chiral multiplet $S_0$
with components $\{z_0,\psi_0, f_0\}$. It also provides the most general
supergravity--matter couplings \FGKVP.

The most general action for chiral matter
supermultiplets $\Sigma$ coupled to supergravity
can then be written as \CFGVP:
\label\superaction
\eq
\Sci=\int d^4x \;\left\{ -{3\over2}
\Dterm{S_0 \S0bar e^{-K_p(\Sigma,\Sigma^*)
/3}}+ \left[\Fterm{S_0^3 W_p(\Sigma)} + \Fterm{{1\over4}f_{ab}(\Sigma)
W^{\alpha a} W^{b}_{\alpha} } + \cc\right]\right\} \, .
\eeq
As usual, the chiral and Weyl weights of the matter chiral multiplets
$\Sigma$ and of the gauge vector multiplet $V$ vanish, while $S_0$ has
unit weights. As in the previous section, we neglect terms with higher
powers of $S_0^{-3}W^{\alpha a} W^b_{\alpha}$, as well as higher derivative
terms.
For future use, the bosonic part
of the component expansion of theory \superaction\ is given in the appendix,
following refs. \CFGVP\ and \KU,
with a discussion on the gauge-fixing of dilatation symmetry which
leads to Poincar\'e supergravity and introduces Newton's constant.
In \superaction, the function $K_p$ is the K\"ahler potential of
the scalar sigma-model if the compensating multiplet
is chosen to normalize canonically the Einstein term.

Notice that the action \superaction\ has an
automatic symmetry under \Kahler\ transformations:
\label\kahlertr
\eq\eqalign{
K_p & \longrightarrow K_p +\varphi(\Sigma) +\varphi^*(
\Sigma^*)\cr
W_p  & \longrightarrow e^{-\varphi(\Sigma)}\,  W_p .}
\eeq
since any such a transformation can be absorbed by redefining $S_0$:
$S_0\longrightarrow e^{\varphi/3}S_0$. K\"ahler invariance
indicates that the action
only depends on the invariant functions ${\cal G}_p = K_p +\log(W_pW_p^*)$
and $f_{ab}$. In conformal supergravity, this information
exhausts the content of K\"ahler symmetry,
which is not to be confused with `active' symmetries like Weyl transformations.

Two symmetries of the superconformal algebra have a particular importance
for us: Weyl and chiral $U(1)$ transformations. These two symmetries
which are not included in the super-Poincar\'e algebra do not commute
with (Poincar\'e) supersymmetry. The chiral $U(1)$ group
is at the origin of the R-symmetry of Poincar\'e theories. Its gauge field
$A_\mu$ is an auxiliary field of minimal Poincar\'e supergravity, as is
the highest component $f_0$ of the compensating multiplet. Weyl
and chiral transformations with parameters $\lambda$ (Weyl)
and $\theta$ (chiral)
act on component fields with a factor
$$
e^{w_j\lambda + in_j\theta/2},
$$
$w_j$ and $n_j$ being the Weyl and chiral weights of the component field.
For a left-handed chiral multiplet $(z,\psi,f)$, one finds the following
weights:
\label\weights
\eq\eqalign{
z :\qquad& w, \hskip1.31cm n=w \qquad\qquad(w : {\rm arbitrary}) , \cr
\psi :\qquad& w+{1\over2}, \qquad n-{3\over2}, \cr
f :\qquad& w+1, \qquad n-3.}
\eeq
Since we are considering chiral matter multiplets $\Sigma$ with $w=n=0$
and since the chiral multiplet of gauge field strength $W^a$ has $w=n=3/2$,
one deduces that the $U(1)$ transformations of (left-handed)
gauginos and chiral fermions are
$$
\lambda^a \longrightarrow e^{3i\theta/4}\lambda^a ,\qquad\qquad
\psi \longrightarrow e^{-3i\theta/4}\psi.
$$
These transformations generate a gauge-chiral
$U(1)$ mixed anomaly clearly controlled by the coefficient
\label\anomcoeff
\eq
c =  {3 \over 16 \pi^2 } \left[ T(G) - \sum_\ssi T\left(R_\ssi \right)
\right],
\eeq
as already discussed in refs. \L, \DFKZa\ and \KL.
It is crucial that these
superconformal symmetries must really be quantum symmetries, and so any
anomalies in them must be cancelled by an appropriate `Green-Schwarz'
counterterm. Such a counterterm can easily be constructed using the chiral
compensating multiplet $S_0$, with $w=n=1$ \DFKZa, \DFKZb, \KL:
\label\cancel
\eq
\Delta \Sci= -2 c \left\{\int d^4x \;\Fterm{
{1\over4}\,\Tr \WW\, \log S_0 }+\cc\, \right\} .
\eeq
This term is claimed to cancel the anomaly to all orders in perturbation
theory \KL. Notice also that the constant $c$
coincides with the coefficient $b$ defined
in eq.~\eqeighteen\ for the case of no charged matter, that we are
considering. This counterterm plays an important role
in what follows.

We now apply the symmetry argument of the global-supersymmetry case.
This argument must be modified in two ways. First, the $R$ and scale
invariances of the global case are now both already contained in the
superconformal invariance, so their consequences follow automatically
from the expression for the superconformal action. Secondly, due to the
presence of the $S_0$ field, and of the anomaly-cancelling Green-Schwarz
term, we need not postulate a non-trivial transformation rule for
$S$ to cancel the scale and $R$ anomalies.

\subsection{Symmetry Arguments}

We start with the Wilson action, for the case of a single condensing  gauge
group. Combining eqs.~\superaction\ and \cancel, the total effective action
can be written as
\label\uno
\eq
\Gamma[S_0,S,\Sigma]=\Gamma_p[S_0,S,\Sigma]+
\Gamma_\ssv[S_0,S]
\eeq
where
\label\dos
\eq
\Gamma_p=\int d^4 x \left\{-{3 \over 2} \Dterm{S_0 S_0^*
\exp\left(-{ K_p(S,S^*,\Sigma,\Sigma^*)\over 3}\right)}
+\left[ \Fterm{S_0^3 W_p(\Sigma)}
+\cc\right]\right\}
\eeq
and $\Gamma_\ssv$ is given (compare with eq.~\whatwewant) by
\label\eqelevent
\eq
\exp \Bigl\{ i\Gamma_\ssv[S,S_0] \Bigr\} = \int \Scd V \;
 \exp \left\{ i \int d^4x \; {1\over 4}
\left[ \Fterm{ \left(S-2c\log S_0\right) \Tr \WW } + \cc \right]  \right\}.
\eeq
This expression automatically incorporates the cancellation of the local Weyl
and $R$ anomalies, and so contains all of the information that was used in
the globally-supersym\-me\-tric case. Since the result is therefore
superconformally invariant, and local, its
leading contribution can also be put into the form of
eq.~\superaction:
\label\newsuperaction
\eq
\Gamma[S,\Sigma,S_0] =\int d^4x \; \left\{ -{3\over2}\Dterm{ S_0 \S0bar
\exp\left[ - \, {K(S,S^*,\Sigma,\Sigma^*) \over 3} \right] } + \left[\Fterm{
S_0^3
W(S,\Sigma) } + \cc\right] \right\} \, .
\eeq

Now comes the main argument for the supergravity case. Inspection of
eq.~\eqelevent\ shows that $\Gamma_\ssv[S,S_0]$ depends on its two
arguments only through the combination $e^{-S/2c} \, S_0$. Since
eq.~\newsuperaction\ completely fixes the result's $S_0$-dependence, we
can immediately read off the nonperturbative super- and \Kahler-potentials:
\label\sugrawilsonresult
\eq \eqalign{
W(S,\Sigma) &= W_p(\Sigma)+W_{np}(S) , \cr
W_{np}(S) &= w \exp\left[ -\ {3S \over 2 c} \right] , \cr
\exp\left(- \; {K(S,\Sbar,\Sigma,\Sigma^*) \over 3} \right) &=
\exp\left(-\; {K_p \over 3} \right) -k \exp\left( -\; {S+\Sbar \over 2c}
\right)
 , \cr}
\eeq
with $k$ and $w$ arbitrary constants.

With these terms, the effective action becomes
\label\combinedresults
\eq\eqalign{
\Gamma[S,\Sigma,S_0] =\int d^4x\, & \left\{ -{3\over 2}\Dterm{S_0S^*_0
\left[ \exp\left( - \; {K_p(S,S^*,\Sigma,\Sigma^*)\over 3} \right) -
k\exp(-\; {S+S^*\over 2c}) \right] } \right. \cr
&  +\left[\Fterm{S_0^3\left(W_p(\Sigma) +we^{-{3S\over 2c}}\right)} + \cc
\right]\bigg\}. } \eeq %

The extension of these results to the case with several condensing factors
in a gauge group is straightforward.  The results are a simple sum of the
corresponding results for a single condensing gauge group:
\label\severalcombinedresults
\eq \eqalign{
W &= W_p+ \sum_n w_n e^{-(3 f_n S / 2 c_n) }, \cr
e^{ - K / 3} &=  e^{-K_p/3} - \sum_n k_n e^{-
(f_n/2 c_n) ( S + \Sbar ) } . \cr}
\eeq
These equations neglect all subleading terms in powers of supercovariant
derivatives.

The neglected higher supercovariant-derivative terms involve the superconformal
generalization of the superfield $\overline{DD}S^*$ of global supersymmetry,
which uses the  `kinetic multiplet' $T(S_0^*S^*)$ \FPvN, \PvN. The construction
of this  multiplet is as follows. Consider a chiral multiplet $\phi$ with
weights
$w=n=1$, and the action
\label\tres
\eq
\int d^4x\, [\phi\phi^*]_D =
\int d^4x\, e[2ff^*-2(D_\mu^c z)(D^\mu_c z^*) +{1\over3}zz^*R] +\ldots,
\eeq
omitting fermionic contributions. Define then the kinetic multiplet by
the identity
\label\cuatro
\eq
\int d^4x\, [\phi\phi^*]_D = \int d^4x\, [\phi T(\phi^*)]_F
= \int d^4x\, e[z f_T + f z_T] + \ldots,
\eeq
where the chiral $T(\phi^*)$ has as components: $(z_T,\psi_T,f_T)$. After a
partial integration, the comparison of \tres\ and \cuatro\ leads to
$z_T =2f^*$ and $f_T = 2\,\Square\,_c z^*= 2\,\Square\, z^*
+{1\over3}Rz^* +\ldots$. The appearance of $T(\phi^*)$ in the
F-density formula \cuatro\ indicates that the kinetic
multiplet has weights $w=n=2$. In our case, we have to consider
$T(S_0^*S^*)$, the kinetic multiplet of $S_0^*S^*$ with $w=-n=1$,
and $\Gamma_V[S,S_0]$ will generically depend on the weight-zero
chiral multiplet
\label\cinco
\eq
S_0^{-2} \, T(S_0^* S^*),
\eeq
or, more generally, on $S_0^{-2}T(S_0^* g(S^*))$, with an arbitrary function
$g$. The factor $S_0^{-2}$ indicates that the components of this multiplet are
suppressed in the super-Poincar\'e theory by one power of the  gauge-fixing
scale of $z_0$, which is the Planck scale.

Therefore, neglected terms involve the dependence of the D-density on the
invariant chiral multiplets with zero weight
$$
S_0^{-2}\,e^{S/c}\,T\big( S_0^* e^{-S^*/2c}\big),
$$
which are suppressed by an additional power of $S_0$.

In order to show the equivalence of the globally supersymmetric limit of
\combinedresults\ with the results obtained in the preceding section, we first
need to discuss the gauge-fixing procedure applied to the compensating
multiplet which leads to the Poincar\'e theory and introduces the Planck
scale. This will be done at the end of this section.

\subsection{The 2PI Effective Action}
\ref\Taylort{T. R. Taylor, \plb{252}{90}{59}.}

We may similarly reproduce the 2PI analysis using local supersymmetry.
We start with a single gauge-group factor, and write the expression for
the generating functional for correlations of gaugino bilinears:
\label\eqtwentytwot
\eq \eqalign{
\exp \Bigl\{i \Wh_{np}[J,S,S_0] \Bigr\} &= \int \Scd V \;
\exp \left\{ i \int d^4x\; {1\over4}\left[ \Fterm{ \left(S-2c \log S_0\right)
\Tr \WW } \right.\right. \cr
& \qquad \qquad \left. + \Fterm{J \Tr \WW} +  \cc \right]  \bigg\}. \cr}
\eeq
This expression includes the anomaly-cancelling term \cancel. Writing the
Legendre transform variable as $\Uhat = \Avg{\Tr \WW}_\ssj$ --- where the
`caret' is introduced for later notational convenience --- we may construct the
2PI effective action, $\Gamma_{np}[\Uhat,S,S_0]$. Since $\Scw_{np}$ depends on
its three arguments only through the one combination $J+S-2c\log S_0$, it
follows
that $$ {\displaystyle {\delta\Gamma_{np}[\Uhat,S,S_0] \over
\delta(S-2c\log S_0)} ={1\over4}\,\Uhat },
$$
and so the 2PI action can be written as
\label\siete
\eq
\Gamma_{np}[\Uhat,S,S_0]=\Xi[\Uhat]+{1\over4}\int d^4 x \left[
\Fterm{\Uhat(S-2c\log S_0)}+\cc \right].
\eeq
Since the second term has anomalous scale and chiral transformations,
these must be cancelled by $\Xi[\Uhat]$, which must therefore include an
anomaly-cancelling term constructed using only the chiral multiplet $\Uhat$,
with weights $w=n=3$. One can then write
\label\ocho
\eq
\eqalign{
\Xi[\Uhat] =& -{3\over2}\int d^4x\, \Dterm{ -a(\Uhat\Uhat^*)^{1/3}}
+\zeta'\int d^4x\,\left[\Fterm \Uhat +\cc \right]\cr
&+{1\over4}\int d^4x\, \left[\Fterm{{2c\over3}\Uhat\log \Uhat }+\cc\right],}
\eeq
with constants $\zeta'$ and $a$. The first two terms are the unique invariant
$D$
and $F$  densities one can write with $\Uhat$ only, and the last
term cancels the anomaly. In other words,
\eq
\eqalign{
\Gamma_{np}[\Uhat,S,S_0] =& -{3\over2}\int d^4x\, \Dterm{
-a(\Uhat\Uhat^*)^{1/3}} \cr
& + {1\over4}\int d^4x\,\left[\Fterm{ \Uhat\left\{ S
+{2c\over3} \log \left( \zeta \Uhat S_0^{-3}\right)\right\}} +\cc\right],}
\eeq
the constant $\zeta$ replacing $\zeta'$. To derive the Poincar\'e theory and
obtain
its K\"ahler potential, it is useful and convenient to work with zero-weight
chiral matter. We then define a new chiral multiplet, $U\equiv
\widehat U S_0^{-3}$, with $w=n=0$, and rewrite
\label\nonpert
\eq
\eqalign{
\Gamma_{np}[U,S,S_0] =& -{3\over2}\int d^4x\, \Dterm{-a S_0S_0^*(UU^*)^{1/3}}
\cr
&+ {1\over4}\int d^4x\,\left[\Fterm{ S_0^3 U\left\{ S
+{2c\over3} \log \left( \zeta U\right)\right\}} +\cc\right].}
\eeq
Comparing with eq. \superaction\ and keeping in mind that
$\Gamma[U,S,S_0] = \Gamma_p +\Gamma_{np}$, we can read
the total K\"ahler \foot\pepito{Notice that different expressions for the
$\ss U$-field K\"ahler potential have been used in the literature for the
supergravity case, without any real derivation. Here we have proved the
uniqueness of expression \ocho, which agrees with the one given in
ref. \Taylort.} and superpotentials,
\label\accpp
\eq\eqalign{
& \exp\left( - \; {K(U,U^*,S,S^*,\Sigma,\Sigma^*)\over 3} \right)  =
\exp\left(-\; {K_p \over 3}\right) -a (U^* U)^{1/3}  , \cr
& W(S,U,\Sigma) =W_p+W_{np};\qquad W_{np}={1\over4}U \left[ S + {2c\over3}
\log \left( \zeta U \right)\right] . \cr}
\eeq

The general 2PI effective action for the case of several condensates
is similarly found to be
\label\eqthirtysevenpp
\eq\eqalign{
\Gamma[U_n,S_0,S,\Sigma]& =  \int d^4x \; \left\{
\Dterm{ S_0 \S0bar \left( \exp \left( -\ {K_p(S,\Sbar,\Sigma,\Sigma^*) \over 3}
\right) - \sum_n a_n (U_n \Ubar_n)^{1/3} \right) } \right. \cr
 &  + \left[\left. \Fterm{ S_0^3 \left(W_p+{1\over4}
\sum_n\,  {2 c_n \over 3} \, U_n \;
\log \left(  \zeta_n  U_n e^{3 f_n S/2c_n}  \right)\right) }
+ \cc\, \right]    \right\} ,}
\eeq
neglecting terms of higher order in $S_0^{-1}$ which
involve in particular the kinetic multiplet.

\subsection{Eliminating the Field $U$}

We may now proceed to eliminate $U$ from the 2PI action, in order to
retrieve our previous result for $\Gamma_\ssv[S,S_0]$.  To do so, we
must solve the equations of motion for the components of the {\it classical}
multiplet $U$, and substitute the result back into $\Gamma_{np}[U,S,S_0]$.

\ref\dqq{J.-P. Derendinger, F. Quevedo and M. Quir\'os,
\npb{428}{94}{282}.}

In the supergravity case, we prefer to begin with a treatment in terms of
components, by using the scalar part of the 2PI action. The component
expansion of the bosonic part of the 2PI effective lagrangian can be derived
from
the general expressions given in the appendix, or from refs. \CFGVP,
\KU\ and \dqq, among others.
Since we are interested in eliminating the multiplet $U$, it
suffices to focus on the nonperturbative part of the action, in which
$U$ appears. With eq.~\nonpert, the terms which depend on $f_0$, $f_u$ or $u$
are given by
\label\auxilloc
\eq
\eqalign{
e^{-1}{\cal L}_{aux}
&={a\over3}(z_0z_0^*)(uu^*)^{-2/3} f_u f_u^*
- 3\Phi f_0 f_0^* + a(uu^*)^{-2/3}[z_0 u^*f_0^*f_u + z_0^*uf_0f_u^*]
\cr
& \qquad \qquad  +\left[z_0^3 f_u (W_{np})_{,u} +z_0^3 f_s
(W_{np})_{,s} + 3 z_0^2 Wf_0 \right] +\cc ,}
\eeq
where $\Phi = e^{-K_p/3}-a(uu^*)^{1/3}$, the superpotential is
$W=W_p +W_{np}(s,u)$, and
\eq
\eqalign{
(W_{np})_{,u} =&\;  {\partial\over\partial u} W_{np}(u,s) =
{1\over4}\left[ s+{2c\over3}\big\{1+\log( \zeta u)\big\}\right], \cr
(W_{np})_{,s} =&\;  {\partial\over\partial s} W_{np}(u,s) =\;
{1\over4}u. }
\eeq
Extremizing with respect to $f_u$, we obtain
\label\auxresult
\eq
\ov{z_0}\ov{f_u} = -\; {3\over a} (uu^*)^{2/ 3} z_0^2 (W_{np})_{,u} - 3 \ov{u}
\ov{f_0},
\eeq
which when put back into eq.~\auxilloc\ gives
\label\auxtwoloc
\eq
\eqalign{
{\cal L}_{aux}=& -{3\over a} \left| z_0^2\, u^{2/3}\,
\W_\cu \right|^2+ {1\over4}z_0^2 u \left( z_0 f_s - 2c f_0 \right) +
\cc \cr
& -3 e^{-K_p/3} f_0f_0^*. }
\eeq
The last term does not depend on $u$. As in the global case, one cannot
solve for the scalar field $u$ (even for constant configurations) without
first eliminating the auxiliary fields $f_s$ and $f_0$, which depend on
$K_p$ and $W_p$. We then proceed as in the global case:
we solve for $u=\widetilde{u}(s,f_s,z_0,f_0)$ as a power series in the
auxiliary fields, with the result this time turning out to be a series in the
combination $\xi \equiv z_0 f_s-2c f_0$.  The first three terms in the
expansion, $\widetilde u=
\widetilde{u}_0 \left(1+\sum_{k=1}^{\infty}\widetilde{u}_k\right)$, which are
the local
version of eq.~\firstfew, are
\label\sugraeff
\eq\eqalign{
\widetilde{u}_0 &= {1\over \zeta e}e^{-3s/2c} =
-\ {3 \over 2c} \; W_{np}(s) , \cr
z_0^2 \widetilde{u}_0\widetilde{u}_1
&=-\ {3a\over 4c^2}\left(\widetilde{u}^*_0\widetilde{u}_0
\right)^{1/3} \xi^* \, \cr
\left(\widetilde{u}^*_0\widetilde{u}^2_0 \right)\widetilde{u}_2 &=
-\ {1 \over 6} \widetilde{u}_0\widetilde{u}_1^*\widetilde{u}_1^*
-\ {1 \over 3}\widetilde{u}_0^* \widetilde{u}_1 \widetilde{u}_1^*, \cr}
\eeq
where $W_{np}(s)$ is defined in eq.~\sugrawilsonresult. The first term in the
expansion, $\widetilde{u}_0$, corresponds to $\xi = 0$ (i.e. it corresponds to
minimizing the  potential to ${\cal O}(\xi^0)$), and is the solution to
$\W_\cu=0$.  The second term, $\widetilde{u}_1$, corresponds to minimizing the
potential to linear order in $\xi$, and so on. Keeping now $\widetilde{u}_0$
and
$\widetilde{u}_1$ from eq.~\sugraeff\ we obtain in ${\cal L}_{aux}$ all
terms quadratic in the auxiliary fields $\xi$. This reproduces
the Wilson action up to higher-derivative terms. Notice, however, that in
this case the leading condition, $\W_\cu=0$, is {\it not} equivalent to
the statement that the strongly-coupled sector does not break
supersymmetry,  which was the standard argument
used in Refs.~\LT,\buddies,\KL, see also Ref.~\LNN for a recent
discussion.
This is because for local supersymmetry, it is
the \Kahler\ derivative, $D_u W_{np} \equiv \W_\cu + K_\cu W_{np}
= W_{np}{\cal G}_u$, rather than
$\W_\cu$, that is the order parameter for supersymmetry breaking.
Nevertheless, it is the solution to $\W_\cu=0$ that provides the leading
stationary point for the 2PI action, and which also reproduces the Wilson
action. This emphasizes the fallacy of using supersymmetry preservation of
the strongly-coupled sector as the argument for determining how to
eliminate $U$.

To eliminate $U$ at the supermultiplet level, use the kinetic multiplet to
rewrite
$$
\Gamma_{np}[U,S,S_0] = {3\over2}a \int d^4x\,
\Fterm{ S_0 U^{1/3} \,T(S_0^*{U^*}^{1/3}) }
+ \left[ \int d^4x\,\Fterm{ S_0^3W_{np}(S,U)} +\cc \right].
$$
The equation determining $U$ is then
\label\localeom
\eq
{a\over2} S_0^{-2}U^{-2/3} T(S_0^*{U^*}^{1/3}) =-\; {\partial\over\partial U}
W_{np}(S,U) = -\; {1\over4}\left[ S+{2c\over3}\big\{1+\log( \zeta
U)\big\}\right],
\eeq
an equation to be interpreted with the superconformal tensor calculus, and
applied to all components of the chiral multiplets appearing in it. This
equation of motion is the local, superconformal generalization of eq. \euler\
derived in the global case. Since the lowest component of $T(S_0^*{U^*}^{1/3})$
is  $2(u^{1/3}f_0 + {1\over3}z_0 u^{-2/3}f_u)^*$, the lowest component of this
equation of motion is again eq. \auxresult,  the equation for the auxiliary
field $f_u$.  As in the global case, one could iteratively derive a solution to
eq.  \localeom\ of the form
$U=\widetilde U=\widetilde U_0(1+\widetilde U_1+\widetilde U_2+\ldots)$, with
$\left.{\partial\over\partial U}W_{np}\right|_{U=\widetilde U_0}=0$, or
\eq
\widetilde U_0 = {1\over \zeta e} e^{-3S/2c},
\eeq
as in the first eq. \firstfew, which is the supermultiplet extension
of the first eq. \sugraeff. It is however important to keep in mind that
equation of motion \localeom\ is derived at the superconformal level. We
are interested in the super-Poincar\'e theory in which the compensator
$S_0$ is not an independent propagating multiplet.

\subsection{The Poincar\'e theory and its global limit}

One of the gauge-fixing conditions imposed to reduce superconformal symmetry
down to super-Poincar\'e is applied on the scalar
component $z_0$ of the compensating chiral multiplet $S_0$. The
microscopic theory with action $\Gamma_p$
contains Einstein terms of the form
$$
-{1\over2} \left(z_0z_0^* e^{-K_p/3}\right) eR,
$$
and a natural choice is
\label\fixmicro
\eq
z_0z_0^* = {1\over\kappa^2} \; e^{K_p/3},
\eeq
(where $\kappa$ is defined in eq.~(A8))
leading to a canonically normalized gravitational lagrangian, with a
field-independent Newton's constant. With this choice, obviously, the $z_0$
contributions in the 2PI effective action $\Gamma_p + \Gamma_{np}$
do not depend on $u$. The same holds for the fermionic component $\psi_0$
of $S_0$, which is constrained by a gauge-fixing condition for special
supersymmetry. The analysis of the elimination of the components of $U$
given in the preceding paragraph applies then to both Poincar\'e and conformal
supergravities.

Using the compensator fixing \fixmicro\ in the 2PI effective action has a
drawback: nonperturbative contributions in $\Gamma_{np}$ also include
gravitational contributions so that the complete Einstein term, which is
\label\Einsteinnp
\eq
-{1\over2\kappa^2}\, \left(1 -a (uu^*)^{1/3}e^{K_p/3}\right)\,eR,
\eeq
is not canonical. As explained in the appendix, a by-product of
non-canonical Einstein terms is the fact that scalar fields are
not in a K\"ahler basis.
Scalar kinetic terms have an additional contribution of the form
\label\noncal
\eq
{3\over4}\kappa^{-2}\left(1 -a (uu^*)^{1/3}e^{K_p/3}\right)
\left( \partial_\mu \log\,[ 1-a(uu^*)^{1/3}e^{K_p/3} ]\right)^2.
\eeq
One can return to the K\"ahler basis by performing a rescaling of the
vierbein of the form [see the appendix]:
$$
e_{m\mu} \quad\longrightarrow\quad [1-a(uu^*)^{1/3}e^{K_p/3}]\, e_{m\mu},
$$
which also redefines the field-dependent Newton constant present in
expression \Einsteinnp.
The resulting theory can be directly obtained
by choosing instead of \fixmicro\ the compensator in such a way
that the Einstein term in the 2PI effective action is canonical.
In the superconformal 2PI action, this term is
$$
-{1\over2}z_0z_0^* \left[ e^{-K_p/3} - a(uu^*)^{1/3}\right] eR\,.
$$
Taking then
\label\newcomp
\eq
z_0 = z_0^* = {1\over\kappa} \left[ e^{-K_p/3} - a(uu^*)^{1/3}\right]^{-1/2}
\eeq
leads to a canonical gravity lagrangian. The `effective' K\"ahler
potential is given by
\label\Keff
\eq
K = -3\log \left[ e^{-K_p/3} - a(uu^*)^{1/3}\right]
= K_p -3\log\left[ 1-a(uu^*)^{1/3} e^{K_p/3} \right].
\eeq

To discuss the global supersymmetry limit of the Poincar\'e supergravity
defined by the K\"ahler potential \Keff\ and the superpotential
$W=W_p+W_{np}$, the first step is to introduce the physical dimensions of
the scalar fields in the theory. The appropriate substitutions are
$$
\eqalign{
K\,\,,\,\, K_p \qquad&\longrightarrow\qquad \kappa^2 K
\,\,,\,\,\kappa^2 K_p \cr
W \qquad&\longrightarrow\qquad \kappa^3 W, \cr
u \qquad&\longrightarrow\qquad \kappa^3 u. }
$$
It is at this point useful to reintroduce the physical dimension of the scalar
field $u$, which describes the gaugino bilinear condensate.
With these rescalings, eq. \Keff\ becomes
$$
\eqalign{
\kappa^2 K =& \kappa^2 K_p -3\log\,\left[1-\kappa^2 a(uu^*)^{1/3}e^{
\kappa^2 K_p/3} \right] \cr
=& \kappa^2\left[ K_p + 3a(uu^*)^{1/3}\right] + {\cal O}(\kappa^4),}
$$
in the flat limit $\kappa\longrightarrow 0$. This result indicates
that the global supersymmetry limit
of the effective 2PI Poincar\'e supergravity has K\"ahler potential
$$
K_{flat} = K_p +3a(uu^*)^{1/3},
$$
as already demonstrated in the previous section.

\section{Moduli couplings and duality anomalies}

\ref\LCO{ J. Lopes-Cardoso and B. Ovrut,
 \npb{369}{92}{351}; \npb{392}{93}{315}.}
\ref\DKL{L. Dixon, V. Kaplunovsky and J. Louis \npb{355}{91}{649}.}
\ref\classic{E. Witten, \plb{155}{85}{151}; \bk
J.-P. Derendinger, L. E. Ib\'a\~nez and H. P. Nilles, \npb{267}{86}{365};\bk
C.P. Burgess, A. Font and F. Quevedo, \npb{272}{86}{661}.}

The main target of application for the above expressions is to the
low-energy limit of (2,2) compactifications of the heterotic string. For
this particular case, there are typically many supermultiplets, $T_\ssa$,
whose scalar components parameterize the moduli space of the string
vacuum being considered, in addition to the model-independent dilaton
supermultiplet, $S$. An important feature of these compactifications are
the target-space symmetries which they exhibit. For many
compactifications the fields $T_\ssa$ transform nontrivially under a
target space duality symmetry group $\Scg$. This transformation is a
symmetry in the sense that it changes the \Kahler\ potential and
superpotential, but it does so only by a \Kahler\ transformation, as in
eq.~\kahlertr. These `duality' symmetries are subject to anomalies which
can also be cancelled by local counterterms \DFKZa, \LCO. Once string loop
effects are included, the moduli fields also modify the gauge couplings due
to threshold corrections \DKL. These corrections can change the tree-level
nonminimal gauge kinetic function, $f_{\rm tree} = S$, by adding to it a
moduli-dependent one-loop contribution: $f=S+\Delta(T)$.  In this case the
role played by $S$ in the previous sections, is instead played by this full
gauge kinetic function $f(S,T)$ \FILQ.

For `realistic' scenarios, for which several factors of a hidden sector gauge
group condense, it has not been clear how to formally
derive  an expression for the
low-energy theory which manifestly displays these symmetries after
supersymmetry breaking. The purpose of this section is to provide
explicit expressions for the low-energy theory, and to show in particular
how the symmetries are all realized in the result. All this as
an application of the discussion of the previous sections.

Let us, for simplicity, discuss the case of an overall modulus field $T$, with
a target-space duality group $\Scg = SL(2,\IZ)$ acting as
\label\modul
\eq
T\longrightarrow{\alpha T-i\,\beta\over i\, \gamma\, T+\delta};
\qquad \hbox{with} \qquad \alpha, \beta, \gamma, \delta \in \IZ,
\qquad \hbox{and} \qquad \alpha\, \delta-\beta\, \gamma=1 .
\eeq
At string tree level, the perturbative part of the low-energy action takes
the form \classic:
\label\treeterms
\eq
K_p^{\rm tree}=-3\log (T+\ov{T})-\log (S+\Sbar), \qquad W_p = 0.
\eeq
This is invariant with respect to the $SL(2,\IZ)$ transformations of
eq.~\modul, with $K_p$ transforming into itself up to a \Kahler\
transformation
\label\kahlertrrr
\eq\eqalign{
K & \longrightarrow K+\varphi(T) + \varphi^*(T^*) \cr
W & \longrightarrow e^{-\varphi(T)}\,  W,\cr
S_0& \longrightarrow e^{{\varphi(T)\over 3}}\, S_0}
\eeq
having $\varphi(T)= 3\log (i\, \gamma\, T +\delta)$ as its parameter.
Notice that the dilaton field, $S$, is invariant under this transformation.

After one-loop string corrections are included, $S$ cannot remain
invariant, however. This has been observed in ref.~\DFKZa, where it was
found that the \Kahler\ potential acquires loop corrections given
by
\label\koneloop
\eq
K_p^{\rm 1-loop} =-3\log \left(T+\ov{T}\right)-\log
 \left[S+\Sbar+3\delta_{\rm GS}\log (T+\ov{T})\right],
\eeq
while the gauge kinetic terms become modified from eq.~\eqelevent\
to
\label\eqeleventt
\eq
\sum_n \Fterm{ \left(f_n S-2c_n\log S_0 +
\left( 2c_n-3f_n\delta_{GS}\right) \log
\eta^2(T) \right) \Tr \WWn }  + \cc .
\eeq
Here $\eta(T)$ is the Dedekind $\eta$ function and the coefficients
$\delta_{GS}$ are explicitly known for the $Z_N$ orbifolds \DFKZa.
These interactions are only invariant under \Kahler\ transformations if the
field $S$ transforms in the following way:
\label\stransf
\eq
S\longrightarrow S+3\delta_{GS}\, \varphi(T) .
\eeq

With this information, we can apply our previous analysis to find
the 2PI effective action, $\Gamma[U_n,S,S_0]$, and
Wilson action, $\Gamma_\ssv[S,S_0]$, that
are induced by gaugino condensation.  For the 2PI effective action we find
the following results for the total \Kahler- and super-potentials:
\label\totalk
\eq\eqalign{
K(S,\Sbar,T,\ov{T}, U_n,\Ubar_n) & =
-3\log\left[\left(S+\Sbar+3\delta_{GS}\log\left(
T+\ov{T}\right)\right)^{1\over 3}\, \left(T+\ov{T}
\right)+e^{-K_2/3}\right. \cr
 & - \left.\sum_n a_n\left(U_n\Ubar_n\right)^{1\over 3} \right],}
\eeq
where $K_2(S,\Sbar,T,\ov{T})$ stands for the (at present unknown)
higher-loop corrections to the \Kahler potential, and
\label\totsup
\eq
W_{np}(S,T,U) ={1\over4}\sum_n U_n\left(f_n S+{2c_n\over3}\log (\xi_n U_n)
+h_n(T) \right),
\eeq
with
\label\hndef
\eq
h_n(T)\equiv\left(2c_n-3f_n\delta_{GS}\right)
\log\eta^2(T)\; .
\eeq

These expressions should be the starting point for the discussion of
gaugino condensation. It is remarkable that our ignorance here lies with
the perturbative rather than the nonperturbative part of the \Kahler
potential! Quantitative results therefore become possible for supersymmetry
breaking in any model for which these perturbative contributions can be
computed. The important point here is that the nonperturbative corrections
to $e^{-K/3}$ due to gaugino condensation, are
only functions of $U$, and these are completely under control since they
are independent of the particular vacuum.
The correct procedure to
see if gauginos actually condense, and if they break supersymmetry,
should be using the 2PI action defined by \totalk, \totsup\ and
\hndef.
This, as far as we know has not been pursued yet, in part
because there was no confidence on what \Kahler\ potential should be
considered for the field $U$ (which  we are providing) and also because it is
simpler just to
work with the Wilson action approach,  which we consider next.

We can obtain the Wilson action below all condensation scales by
eliminating the superfields $U_n$. We can follow a procedure similar
to that of section 4.3. The extremal for $f_{U_n}$ yields,
\label\cambio
\eq
z_0^* f_{U_n}^*=3 z_0^2 \left|u_n\right|^{4/3}(W_{np})_{,u_n}
-3u_n^* f_0^* .
\eeq
The stationary condition for $u_n$ can be solved by expanding as before
$u_n=\widetilde{u}_{n0}\left(1+\widetilde{u}_{n1}+
\ldots\right)$. One obtains,
\label\seveff
\eq\eqalign{
\widetilde{u}_{n0} &= -\ {3 \over 2c_n} \; w_n
\exp\left\{-\ {3 \left(f_n s+h_n(T)\right) \over 2 c_n} \right\} , \cr
z_0^2 \widetilde{u}_{n0} \widetilde{u}_{n1} &
=-\ {3a_n \over 4c_n^2}\left(\widetilde{u}^*_{n0}\widetilde{u}_{n0}
\right)^{1/3} \xi^*_n \, ,\cr
\left(\widetilde{u}^*_{n0}\widetilde{u}^2_{n0} \right)\widetilde{u}_{n2} &=
-\ {1 \over 6} \widetilde{u}_{n0}\widetilde{u}_{n1}^*\widetilde{u}_{n1}^* -\ {1
\over 3}
\widetilde{u}_{n0}^* \widetilde{u}_{n1} \widetilde{u}_{n1}^*, \cr}
\eeq
where the auxiliary field around which the expansion is done now reads
\label\expsev
\eq
\xi_n=z_0^2\left(f_s + h_n(T)_{,T}\ f_T\right)-2c_nf_0.
\eeq
{}From the terms $\widetilde{u}_{n0}$ and $\widetilde{u}_{n1}$, we obtain the
Wilson
action with superpotential
\label\superfinal
\eq
W_{np}(S,T)=\sum_n w_n \exp\left\{ -\ {3 \over 2c_n}\ (f_n S+h_n(T))\right\},
\eeq
and K\"ahler potential
\label\kahfinal
\eq\eqalign{
K(S,S^*,T,T^*)&=-3\log\left[
\left(S+S^*+3\delta_{GS}\log(T+T^*)\right)^{1/3}(T+T^*)+e^{-K_2/3}\right. \cr
 &\left. -\sum_n k_n \exp\left\{ -\ {1 \over 2c_n }\left(f_n\left(S+S^*\right)
+h_n(T)+h_n^*(T^*)\right)\right\}\right] . }
\eeq

Notice that the effective action is invariant under \Kahler\
transformations, and so also under $SL(2,\IZ)$ transformations, provided
that the field $U$ transforms under duality as its definition would suggest:
$U\rightarrow e^{-\varphi(T)}\, U$ and the unkown perturbative corrections
$e^{-K_2/3}$ transform properly.

\section{Conclusions}

We have presented here a systematic treatment of the process of gaugino
condensation in supersymmetric $N=1$ Yang-Mills theories coupled to
neutral scalar fields. Our analysis has accomplished several things:

\ref\ILS{K. Intriligator, R.G. Leigh and N. Seiberg, \prd{50}{94}{1092}.}
\topic{1}
It has put some previous approaches on a firmer basis by showing how they
can be better interpreted in terms of the 2PI effective and Wilson actions. In
particular, our modification of the VY approach solves a minor puzzle as to
why the heavy degree of freedom corresponding to a {\it quantum} field, $U$,
can consistently appear in the low energy theory below the condensation scale.
In our approach the problem does not arise because $U$ is a
classical
field corresponding to a composite operator whose expectation we wish to study.
As a result, in order to retrieve the Wilson action from the effective
action for $U$, $U$ must simply be eliminated using its field equations. This
is {\it not} to be regarded as the tree approximation to `integrating it out'
as a quantum field. We regard our treatment to also shed light on the more
recent
treatments in which
$U$ is regarded as a quantum
field which is `integrated in' \ILS.

Another interesting
 discussion of gaugino condensation uses
the Nambu-Jona-Laisinio approach \DMR. In this case $U$  is also treated
as a quantum field and Coleman-Weinberg corrections to its potential
are considered, potentially leading to  modifications of the standard
results of \VY, \DRSW. Although such corrections can and do arise in
a theory containing a quantum field $U$, they are irrelevant in our
approach, since $U$ is classical. Our treatment shows how the general
results of \VY, \DRSW\ (such as the inpossibility of fixing the $vev$ of $S$
with a single condensate) are more robust than would be expected for
an approximate treatment of a theory of a quantum field $U$.
\foot\graham{There are other differences of detail as well.
In \DMR, the field $\ss U$ is identified as a Goldstone boson
for the $\ss R$-symmetry. In our approach the corresponding
Goldstone boson is the axion field $\ss \Im S$, and is not
related to $\ss U$ (which is, after all, only classical).
In the case of several condensates, there the $\ss R$-symmetry is only
approximate and one expects at best to have pseudo-Goldstone bosons,
since the axion gets a mass even if supersymmetry is not broken.
(We thank Graham Ross for helping us clarify these points.)}

\topic{2}
We have further shown, both by general arguments, and by explicit calculation,
how the approaches based on the 2PI effective action and the Wilson action give
equivalent results for the low-energy theory. Furthermore, since the
important issue here is not only to find the effective action below
condensation, but also to trace the condensation process, we can say that the
2PI action formalism is more suitable for the discussion than the Wilson
approach, since using it we can learn if the condensate actually forms or not.
Moreover, to be able to use the Wilson approach we have to know the  light
degrees of freedom beforehand. We have previously encountered a case
\BDQQ\ for which it is not possible to use the Wilson approach directly,
since the degrees of freedom change after the condensation process. In that
case
a massless two-index tensor of the underlying  theory is replaced by a massive
three-index tensor in the effective theory below condensation. This we
discovered by using the 2PI  effective action approach, of course after that
identification and the corresponding elimination of $U$ we can find the Wilson
action below condensation. In this sense we see the 2PI approach as more
fundamental.

\topic{3}
We have obtained new results concerning the \Kahler\ potential of the low
energy effective theory below the condensation scale, in addition to
reproducing earlier work for the superpotential, which we put into
a firmer basis, especially the several condensates case. We found calculable
nonperturbative corrections to the \Kahler\ potential, as well as many
other terms involving higher numbers of supercovariant  derivatives of the
light fields. These higher-derivative terms give corrections only to the
\Kahler\ potential, and are systematically suppressed by inverse powers of
the condensation scale, $1/\Lc$.

\topic{4}
Our work could shed some light on the question of supersymmetry breaking due to
gaugino condensates in superstring  theories. Besides showing how to
write effects of the condensation in a way that manifestly respects the
duality symmetries, our determination of the \Kahler\ potential may require a
reanalysis of the previous phenomenological studies of the low energy
scalar potential, and the soft supersymmetry-breaking terms.
However, presently unknown perturbative corrections to the \Kahler\
potential will be more important, in the weak coupling
regime, than the non-perturbative ones we have found. Further progress
must wait for these corrections to be computed in particular models.

Finally, we mention in passing that our techniques are generalizable to the
case where the condensing system also contains charged matter. Such
models have played an important role in globally supersymmetric theories,
and can arise in the hidden sector of string theories, where it permits
more freedom to obtain minima with supersymmetry broken at a
phenomenologically interesting scale.

\bigskip

\centerline{\bf Acknowledgements}

\bigskip

We would like to thank A. Font, L. Ib\'a\~nez,
R. Myers, S. Peris, M. Peskin, G. Ross, C. Savoy,
G. Veneziano and S. Yankielowicz for  helpful conversations.
This research was partially funded by N.S.E.R.C.\ of Canada, les
Fonds F.C.A.R.\ du Qu\'ebec, the Swiss National Foundation,
the European Union (contracts SC1$^*$-CT92-0789 and CHRX-CT92-0004)
and the CICYT of Spain (contract AEN94-0928). C.B. would
like to thank the Institut de Physique of the Universit\'e de
Neuch\^atel, for its warm hospitality, and for providing such a pleasant and
convivial setting in which to pursue this work.

\bigskip\noindent\vfill\eject
\centerline{\sectionfont \bf Appendix }
\bigskip\rm\nobreak

\nequation = 1

In section 4, we have formulated supergravity actions in the
framework of superconformal
tensor calculus. This appendix enumerates some results which are useful in
component expansions of tensor calculus expressions.

We consider the general supergravity lagrangian for a set of chiral multiplets
$\Sigma^i$, with zero chiral and Weyl weights, and a chiral multiplet
$S_0$, with unit weights, used as compensator. We will restrict ourselves
to the contributions of the bosonic components $z^i,f^i$ and $z_0,f_0$ of
$\Sigma^i$ and $S_0$ respectively.

The general conformal supergravity for these multiplets is
defined by the expression
\clabel\appone A
\eq
{\cal L}=
-\; {3\over2}\left[ S_0 S_0^*\Phi(\Sigma^i, \Sigma_i^*)\right]_D
+ [S_0^3 w(\Sigma^i)]_F +\cc
\eeq
The real function $\Phi$ and the superpotential $w$ are arbitrary.

The bosonic part of this lagrangian density also depends on the bosonic
gauge fields of the superconformal algebra, some of them being algebraic (like
the spin connection) or gauge-fixed to reduce the symmetry and
obtain the Poincar\'e theory (this is the case of the gauge field of scale
transformation). Only the metric tensor and the chiral $U(1)$ gauge field
$A_\mu$ (which is auxiliary)
will participate in the bosonic terms of the Poincar\'e theory.

After the elimination of all auxiliary fields,
a convenient expression for the bosonic part of this theory is
\clabel\apptwo A
\eq
e^{-1}{\cal L}_{bos} = -\; {1\over2}(z_0 z_0^*\Phi) R
+\; {3\over4}(z_0z_0^*\Phi)\left[\partial_\mu{\rm log}(z_0z_0^*\Phi)\right]^2
-(z_0z_0^*\Phi)K^i_j(\partial_\mu z^j)(\partial^\mu z_i^*)
-V_0,
\eeq
with $\Phi=\Phi(z^i,z_i^*)$, $w=w(z^i)$, and with the definitions
\clabel\appthree A
\eq
K=-3\,{\rm log}\Phi, \qquad \Phi= e^{-K/3},\qquad
K^i_j = {\partial^2\over\partial z^j\partial z_i^*}\; K.
\eeq
These scalar kinetic terms are obtained after solving
for the auxiliary $A_\mu$. Defining further
\clabel\appfour A
\eq
{\cal G} = K +{\rm log} (ww^*), \qquad
{\cal G}_i = {\partial\over\partial z^i}\; {\cal G}, \qquad
{\cal G}^i = {\partial\over\partial z_i^*}\; {\cal G},
\eeq
the scalar potential reads
\clabel\appfive A
\eq
V_0= (z_0 z_0^*\Phi)^2 e^{\cal G}\left[ (K^{-1})^i_j\,{\cal G}^j
{\cal G}_i -3 \right].
\eeq
It is generated by the auxiliary scalar field lagrangian
\clabel\appaux A
\eq
\eqalign{
e^{-1}{\cal L}_{aux}
=& -3z_0z_0^*\Phi^i_jf_i^*f^j -3\Phi f_0^*f_0-3z_0\Phi_i f_0^*f^i
-3z_0^*\Phi^if_i^*f_0 \cr
&\qquad +[3z_0^2wf_0+z_0^3w_if^i] +\cc \cr
=& \,\,(z_0 z_0^*\Phi)K^i_j f_i^* f^j
-3\Phi {\widetilde f}^*\widetilde f +[3z_0^2w\widetilde f +z_0^3wf^i{\cal G}_i
] +\cc\,\,,}
\eeq
with $\widetilde f = f_0 +\Phi^{-1}z_0\Phi_if^i$, which implies
\clabel\appf A
\eq
f^i = -(z_0 z_0^*\Phi)^{-1} {z_0^*}^3  w^*
(K^{-1})^i_j{\cal G}^j, \qquad\qquad
\widetilde f = \Phi^{-1} {z_0^*}^2w^*.
\eeq

The super-Poincar\'e invariant theory is obtained by gauge-fixing of
the unwanted superconformal symmetries. This procedure includes conditions
of the form
\clabel\compfix A
\eq
z_0 = z_0^* = \kappa^{-1} h(z^i, z_i^*),
\qquad \kappa^{-1} = {M_P\over\sqrt{8\pi}} \simeq 2.4\times 10^{18}\,{\rm GeV},
\eeq
which fixes scale and chiral $U(1)$ symmetries and
eliminate the scalar component of the compensator.

The most natural fixing condition is obtained when imposing that
the Einstein term in the supergravity lagrangian is canonically
normalized:
$$
z_0 z_0^*\Phi = \kappa^{-2},
$$
or $h= \Phi^{-1/2}$. The bosonic lagrangian is then
\clabel\appsix A
\eq
\eqalign{
e^{-1}{\cal L}_{Poin} &= -{1\over2}\kappa^{-2}R
-\kappa^{-2}K^i_j(\partial_\mu z^j)(\partial^\mu z_i^*) -V, \cr
V &= \kappa^{-4} e^{\cal G}\left[ (K^{-1})^i_j\,{\cal G}^j{\cal G}_i
-3\right].  }
\eeq
With this choice of compensator,
the function $K$ is the K\"ahler potential of the Poincar\'e theory,
and the lagrangian only depends on the combination
${\cal G}=K+{\rm log}(ww^*)$ and on its derivatives.

On the other hand, choosing a general gauge-fixing condition \compfix\
leads to
\clabel\appseven A
\eq
\eqalign{
e^{-1}{\cal L}_{Poin}^\prime &= -{1\over2}\kappa^{-2}(h^2\Phi)R
+{3\over4}\kappa^{-2}(h^2\Phi)^{-1}[\partial_\mu(h^2\Phi)]^2
-\kappa^{-2}(h^2\Phi)K^i_j(\partial_\mu z^j)(\partial^\mu z_i^*)
- V^\prime, \cr
V^\prime &= \kappa^{-4}(h^2\Phi)^2 e^{\cal G}\left[
(K^{-1})^i_j{\cal G}^j{\cal G}_i-3\right]. }
\eeq
Since the two lagrangians ${\cal L}_{Poin}$ and
${\cal L}_{Poin}^\prime$ differ in the fixing of dilatation symmetry,
they should be related by a simple rescaling of the vierbein. If:
$$
e_{m\mu} \longrightarrow (h^2\phi)^{-1/2}e_{m\mu}, \qquad
e \longrightarrow (h^2\phi)^{-2} e,
$$
then:
$$
\eqalign{
-{1\over2}(h^2\Phi)eR &\quad\longrightarrow\quad
-{1\over2}eR-{3\over4}e[\partial_\mu{\rm log}(h^2\phi)]^2, \cr
{3\over4}e(h^2\Phi)^{-1}[\partial_\mu(h^2\Phi)]^2 &\quad\longrightarrow\quad
{3\over4}e(h^2\Phi)^{-2}[\partial_\mu(h^2\Phi)]^2 =
{3\over4}e[\partial_\mu{\rm log}(h^2\phi)]^2, \cr
e(h^2\Phi)K^i_j(\partial_\mu z^j)(\partial^\mu z_i^*)
&\quad\longrightarrow\quad
eK^i_j(\partial_\mu z^j)(\partial^\mu z_i^*), \cr
eV^\prime &\quad\longrightarrow\quad e\kappa^{-4}e^{\cal G}\left[
(K^{-1})^i_j{\cal G}^j{\cal G}_i-3\right]=eV, }
$$
and, finally,
$$
{\cal L}_{Poin}^\prime \quad\longrightarrow\quad {\cal L}_{Poin}.
$$

In the Poincar\'e theory \appsix, all fields are dimensionless. Formally,
their canonical dimension can be restored by the rescalings
$$
\eqalign{
K \qquad&\longrightarrow\qquad \kappa^2 K, \cr
W \qquad&\longrightarrow\qquad \kappa^3 W. }
$$
The bosonic lagrangian becomes
\clabel\flatone A
\eq
\eqalign{
e^{-1}{\cal L}_{Poin} =& -{1\over2}\kappa^{-2}R -K^i_j \,(\partial_\mu z^j)
(\partial^\mu z^*_i) - V, \cr
V =& e^{\kappa^2K}\left[ (K^{-1})^i_j(W_i+\kappa^2 WK_i)
({W^*}^j +\kappa^2 W^* K^j) - 3\kappa^2 WW^* \right]. }
\eeq
The flat limit, obtained with $\kappa\longrightarrow 0$, is then
$$
{\cal L}_{flat} = -K^i_j \,(\partial_\mu z^j)(\partial^\mu z^*_i) -
(K^{-1})^i_j\,W_i \,{W^*}^j,
$$
with flat K\"ahler potential $K$ and superpotential $W$.

\listrefs

\bye